\documentclass[twocolumn]{aastex63}
\usepackage{graphicx,natbib,color,bm,times}
\usepackage{amsmath}
\usepackage{mathtools}
\graphicspath{{./fig/}{./png/}}
\newcommand{\EQ}{\begin{equation}}
\newcommand{\EN}{\end{equation}}
\newcommand{\EQL}{\begin{align}}
\newcommand{\ENL}{\end{align}}
\newcommand{\EQA}{\begin{eqnarray}}
\newcommand{\ENA}{\end{eqnarray}}
\newcommand{\eq}[1]{Eq.~(\ref{#1})}

\newcommand{\eqs}[2]{Eqs.~(\ref{#1}) and~(\ref{#2})}
\newcommand{\eqss}[2]{Eqs.~(\ref{#1})--(\ref{#2})}
\newcommand{\Eq}[1]{Equation~(\ref{#1})}
\newcommand{\Eqs}[2]{Equations~(\ref{#1}) and~(\ref{#2})}

\newcommand{\Eqss}[2]{Equations~(\ref{#1})--(\ref{#2})}

\newcommand{\Sec}[1]{Sect.~\ref{#1}}

\newcommand{\Fig}[1]{Figure~\ref{#1}}

\newcommand{\Figs}[2]{Figures~\ref{#1} and \ref{#2}}

\newcommand{\Tab}[1]{Table~\ref{#1}}

{}
{}
\newcommand{\meanFFFF}{\overline{\mbox{\boldmath ${\cal F}$}}{}}{}
\newcommand{\meanemf}{\overline{\cal E} {}}

\newcommand{\meanSSSS}{\overline{\mbox{\boldmath ${\mathsf S}$}} {}}

{}
{}
\newcommand{\meanEMF}{\overline{\mbox{\boldmath ${\cal E}$}}{}}{}
\newcommand{\meanEEEE}{\overline{\mbox{\boldmath ${\cal E}$}}{}}{}
{}{}
{}{}
{}
{}
{}
{}
{}
{}
{}
{}
{}
{}
{}
{}
{}
\newcommand{\meanUU}{\overline{\bm{U}}}

{}

\newcommand{\meanA}{\overline{A}}
\newcommand{\meanB}{\overline{B}}

\newcommand{\meanH}{\overline{H}}
\newcommand{\meanU}{\overline{U}}

\newcommand{\meanJ}{\overline{J}}

\newcommand{\meanQ}{\overline{\cal Q}}
\newcommand{\meanEEE}{\overline{\cal E}}
\newcommand{\meanFFF}{\overline{\cal F}}

{}

{}
{}
{}
{}

\newcommand{\Ot}{\tilde{\Omega}}
\newcommand{\Sti}{\tilde{S}}

\newcommand{\pphi}{\hat{\bm{\phi}}}

\newcommand{\xxx}{\hat{\mbox{\boldmath $x$}} {}}
\newcommand{\yyy}{\hat{\mbox{\boldmath $y$}} {}}
\newcommand{\zzz}{\hat{\mbox{\boldmath $z$}} {}}

\newcommand{\meanAA}{{\overline{\bm{A}}}}
\newcommand{\meanBB}{{\overline{\bm{B}}}}
\newcommand{\meanJJ}{{\overline{\bm{J}}}}

\newcommand{\kk}{\bm{k}}

\newcommand{\aaaa}{\bm{a}}
\newcommand{\jj}{\bm{j}}

\newcommand{\jjB}{\bm{j}^{(B)}}
\newcommand{\jjz}{\bm{j}^{(0)}}
\newcommand{\jjM}{\bm{j}^{(\rm mr)}}

\newcommand{\bb}{\bm{b}}
\newcommand{\bbRob}{\bm{b}_{\rm Rob}}

\newcommand{\bbB}{\bm{b}^{(B)}}

\newcommand{\bbz}{\bm{b}^{(0)}}
\newcommand{\bbM}{\bm{b}^{(\rm mr)}}

\newcommand{\BB}{\bm{B}}

\newcommand{\JJ}{\bm{J}}

\newcommand{\AAA}{\bm{A}}

\newcommand{\UU}{\bm{U}}
\newcommand{\UUS}{\bm{U}^{(S)}}

\newcommand{\FFK}{\bm{F}_{\rm K}}
\newcommand{\FFM}{\bm{F}_{\rm M}}
\newcommand{\ffK}{\bm{f}_{\rm K}}
\newcommand{\ffM}{\bm{f}_{\rm M}}

\newcommand{\uu}{\bm{u}}

\newcommand{\uuz}{\bm{u}^{(0)}}
\newcommand{\uuM}{\bm{u}^{(\rm mr)}}
\newcommand{\hM}{h^{(\rm mr)}}
\newcommand{\aaz}{\bm{a}^{(0)}}
\newcommand{\hz}{h^{(0)}}

\newcommand{\uuB}{\bm{u}^{(B)}}
\newcommand{\aaB}{\aaaa^{(B)}}
\newcommand{\hB}{h^{(B)}}

\newcommand{\ssB}{\boldsymbol{{\mathsf s}}^{(\bm{{B}})}}
\newcommand{\ssz}{\boldsymbol{{\mathsf s}}^{(\bm{{0}})}}

\newcommand{\nab}{{\bm{\nabla}}}

\newcommand{\OO}{\bm{\Omega}}
\newcommand{\ppsi}{\mbox{\boldmath $\psi$} {}}

\newcommand{\ttau}{\mbox{\boldmath $\tau$} {}}

\newcommand{\SSSS}{\mbox{$\boldsymbol{\mathsf S}$} {}}
\newcommand{\ssss}{\mbox{$\boldsymbol{\mathsf s}$} {}}

\newcommand{\ii}{{\rm i}}

\newcommand{\DDD}{{\cal D} {}}

\def\H{\mbox{\rm H}}

\def\ShK{\mbox{\rm Sh}_{\rm K}}

\def\Pm{\mbox{\rm Pr}_{\rm M}}
\def\Rm{\mbox{\rm Re}_{\rm M}}
\def\Rey{\mbox{\rm Re}}

\def\Lu{\mbox{\rm Lu}}

\def\cs{c_{\rm s}}

\def\kf{k_{\rm f}}

\def\kB{k_B}

\def\Brms{B_{\rm rms}}

\def\urms{u_{\rm rms}}
\def\urmst{u_{{\rm rms}}(t)}

\def\brms{b_{\rm rms}}
\def\Brmst{B_{{\rm rms}}(t)}

\def\etat{\eta_{\rm t}}

\def\etaT{\eta_{\rm T}}
\def\eeta{\boldsymbol{\eta}}
\def\aalpha{\boldsymbol{\alpha}}

\def\pphi{\boldsymbol{\phi}}
\def\ppsi{\boldsymbol{\psi}}
\def\ssigma{\boldsymbol{\sigma}}
\def\ttau{\boldsymbol{\tau}}

\newcommand{\W}{\,{\rm W}}
\newcommand{\V}{\,{\rm V}}

\newcommand{\T}{\,{\rm T}}
\newcommand{\G}{\,{\rm G}}

\newcommand{\K}{\,{\rm K}}
\newcommand{\g}{\,{\rm g}}
\newcommand{\s}{\,{\rm s}}

\newcommand{\m}{\,{\rm m}}

\newcommand{\J}{\,{\rm J}}

\newcommand{\A}{\,{\rm A}}

\def\blue{\textcolor{blue}}
\def\apjl{ApJL}
\def\apjs{ApJS}

\def\blue{\textcolor{black}}

\begin{document}
\correspondingauthor{Maarit J. K\"apyl\"a}
\email{maarit.kapyla@aalto.fi}
\author[0000-0002-9614-2200]{Maarit J. K\"apyl\"a}
\affiliation{Department of Computer Science, 
	      Aalto University, P.O. Box 15400, FI-00076 Aalto, Finland}
\affiliation{Max Planck Institute for Solar System Research,
              Justus-von-Liebig-Weg 3, D-37077 G\"ottingen, Germany}	      
\affiliation{Nordita, KTH Royal Institute of Technology and Stockholm University, 
              Hannes Alfv\'ens v\"ag 12, SE-10691 Stockholm, Sweden}
  
\author[0000-0001-9840-5986]{Matthias Rheinhardt}
\affiliation{Department of Computer Science, 
	      Aalto University, PO Box 15400, FI-00076 Aalto, Finland}

\author[0000-0002-7304-021X]{Axel Brandenburg}
\affiliation{Nordita, KTH Royal Institute of Technology and Stockholm University, 
              Hannes Alfv\'ens v\"ag 12, SE-10691 Stockholm, Sweden}
\affiliation{The Oskar Klein Centre, Department of Astronomy,
              Stockholm University, AlbaNova, SE-10691 Stockholm, Sweden}
\affiliation{McWilliams Center for Cosmology \& Department of Physics, Carnegie Mellon University, Pittsburgh, PA 15213, USA}
              
\title{Compressible test-field method and its application to shear dynamos}
\shorttitle{Compressible test-field method and shear dynamos}
\shortauthors{K\"apyl\"a, Rheinhardt \& Brandenburg}

\date{\!$ \, $Revision: 1.348 $ $\!}

\begin{abstract}
In this study we present a compressible test-field method
(CTFM) for computing $\alpha$ effect and turbulent magnetic
diffusivity tensors, as well as those relevant for
mean ponderomotive force and mass source,
applied to the full MHD equations.
We describe the theoretical background of the method,
and compare it
to the quasi-kinematic test-field method, and to
the previously studied variant working in simplified MHD (SMHD).
We present several test cases using
velocity and magnetic fields of the Roberts geometry,
and also compare with the imposed-field method.
\blue{We show that,
for moderate imposed field strengths,
the nonlinear CTFM (nCTFM)    
gives results in agreement with the imposed-field method. 
Comparison of different flavors of the nCTFM in the shear dynamo case
also agree up to equipartition field strengths.
}
Some deviations between the CTFM and SMHD variants exist.
As a relevant physical application, we study non-helically forced
shear flows, which exhibit large-scale dynamo action,
and present a re-analysis of low Reynolds number, moderate
shear systems, where we previously neglected the pressure gradient in the
momentum equation, and found no coherent shear-current effect.
Another key difference is that in the earlier study we used magnetic
forcing to mimic small-scale dynamo action, while here it is
self-consistently driven by purely kinetic forcing.
\blue{The kinematic CTFM
with general validity forms the core of our analysis.}
We still find no coherent shear-current effect, but do recover strong
large-scale dynamo action that, according to our analysis, is
driven through the incoherent effects.
\end{abstract}

\section{Introduction}

\blue{
Over the past few decades, both local and global numerical simulations
of accretion discs have demonstrated that magnetic fields can be
generated 
by a dynamo
and drive turbulent accretion through what is believed to be   
the magneto-rotational instability \citep{BNST95,HGB96,Hawley00}.
Real discs are always stratified about the midplane, which can lead to
kinetic helicity and thereby to an $\alpha$ affect.
Whether or not this really explains what is seen in numerical
simulations is unclear, because there are other, potentially more
powerful alternatives.
One of them is the shear-current (SC) effect.
It is a mean-field dynamo effect that can, in
}
principle, generate large-scale magnetic fields based on the off-diagonal
components of the turbulent magnetic diffusivity
tensor \citep{RK03,RK04}.
Whether or not this effect
can also be responsible for the large-scale dynamo (LSD) seen in some
non-helically forced
shear flows continues to raise debate \citep[see, e.g., the recent
papers][with contradictory results for and against]{SB16,SMHD,ZB21}.
Such a dynamo has been proposed to avoid the need for the more classical
helicity-based $\alpha$ effect in situations where
stratification and rotation are ineffective or absent, and hence neither
kinetic helicity nor $\alpha$ effect can arise.
One major cause of the aforementioned contradictions arises from the
fact that, methodologically, a reliable
quantitative measurement device capable
of returning the turbulent transport coefficients for
an MHD background turbulence due to
a small-scale dynamo (SSD)
in full MHD has not been available.

The imposed-field method was the first machinery 
developed in the 1990's \citep{BTNPS90} for the retrieval of 
$\alpha$  effect and turbulent pumping by 
imposing uniform magnetic fields 
in different directions, measuring the 
mean (a.k.a. turbulent)
electromotive force (EMF), and solving for the unknown
coefficients.
Gradients of the mean field contribute to the 
mean EMF via the turbulent diffusivity tensor, hence
it is important to guarantee that 
they remain weak in the evolving magnetic field.
Therefore, the field has to be reset after suitable time
intervals, and not taking this into account  properly has led to some
misinterpretations of the results \citep[see the discussion in][and the references therein]{OSB01,KKB10}.
When properly used, this method continues to be a valuable tool, and is used also in this work to 
validate the test-field results in the simplest cases.

The next methodology was introduced by \cite{BS02} as a method of moments: 
to tackle the large amount of unknown 
transport coefficients entering the mean EMF
while there are only three equations relating it to the mean field, additional
equations were constructed by forming 
a sufficient number of moments.
Measuring them from the numerical models 
enables then to solve for the
coefficients. This method 
can retrieve both the $\alpha$  
and the turbulent resistivity 
tensors, but relies on the mean field, generated in the system, to be 
unsteady, typically oscillatory.
This method has various incarnations in many astrophysical  contexts \citep[e.g.][]{SB15b,Shi16,Simard16,KGVS18}. 
Occasionally, however, it is employed in an improper way 
by pre-assuming some of the coefficients to be negligible;  putting them
deliberately to zero then
renders the fitting meaningless 
in the worst case
 \citep[e.g.][]{Simard13,SB15b,Shi16}.

The third alternative is the test-field method (TFM), introduced by
\cite{Schrinner05,Schrinner07}, where linearly independent test fields
are subjected to the velocity taken from a simulation,
including both its fluctuating and mean constituents. The test fields 
are passive, i.e., do not affect the course of the simulation itself
(except possibly through the time step control).
The equations for the corresponding fluctuating magnetic fields are
solved, which then allows for retrieving
the full set of tensor coefficients \citep{BRRS08}.
If the simulation is purely hydrodynamic, this approach is 
kinematic,
but if it is an MHD run, where the generated magnetic field
does backreact on the flow, the method is called 
quasi-kinematic (QKTFM).
For both variants, however, the same procedures apply.
This method has proven immensely successful, and has been 
utilized within a broad spectrum of 
astrophysical
stellar, planetary, and disk dynamo applications, 
including the shear dynamo problem \citep{BRRK08}.
\blue{The QKTFM has been applied in Cartesian domains with and without shearing-periodic 
boundary conditions and horizontal ($xy$) averaging,
as well as
in spherical domains, with longitudinal
averaging. 
The latter type of models \cite[see e.g.,][]{Warnecke2017a} has included rotation and stratification,
and is hence not an optimal situation
for clarifying whether or not the SC effect can be important.
Also,
it is not trivial to separate this effect from others contributing to 
the turbulent magnetic diffusivity tensor
such as the R\"adler effect with its antisymmetric contribution.}
When an SSD is excited in such setups, it
is supposed to boost the SC effect \citep[the scenario 
proposed by][]{SB16}.
The (Q)KTFM however, does not apply: 
SSD action generates a magnetic
background turbulence, the magnetic part of which is not accounted for.

A core machinery towards a TFM, which can
take into account the magnetic background turbulence, was presented by \citep{RB10}, 
albeit relying on simplified MHD (SMHD), where 
pressure gradient and self-advection of the flow were dropped from the momentum equation.
Another step further was taken in \cite{SMHD},  
admitting self-advection, while yet ignoring the pressure gradient and hence variations of density.
This study 
has been deemed inconclusive \citep[see, e.g.][]{ZB21}, as 
second-order correlation approximation (SOCA) calculations 
of \cite{SB16} were interpreted to indicate a decisive role of the pressure
gradient in creating the magnetic SC effect.
In contrast, we argued that the magnetic SC effect
continues to exist in SMHD in the
ideal limit, albeit with a sign not supportive for a mean-field dynamo.
This paper aims at introducing a method meeting all requirements,
namely the fully 
\blue{compressible} 
test-field method (CTFM).
We present test cases to prove its functionality, and then make a first
attempt to apply it to the shear dynamo problem in the regime of moderate
Reynolds number, 
\blue{magnetic Prandtl number, and}
shear, where we measure the turbulent transport
coefficients and interpret them in the framework of mean-field dynamo theory.

\blue{
We should clarify from the outset that our primary goal in the calculation
of mean-field transport coefficients is 
at present not to utilize them in more
economic mean-field models of astrophysical dynamos, 
but rather to provide some understanding of the turbulent processes
found in the progressively more realistic simulations
of such dynamos.
In principle, our computations can also identify specific targets of what
to look for in future studies.
We further
note that for the first steps to understand the SC effect with the CTFM,
we use the simplest possible setup, excluding 
physically important effects
such as rotation and stratification, 
thus studying 
this effect in isolation. This approach is chosen to 
gain physical insights about this effect, albeit   
not allowing us to
assess its relevance in astrophysical objects.
}

\section{Model and Methods}

The full MHD (FMHD) system of equations,
here with an isothermal equation of state, is more complex than 
that of SMHD
used in earlier TFMs because
of the occurrence of the pressure gradient.
As a result, 
we need
an additional evolution equation for the density,
being the counterpart to the Poisson equation for the pressure in incompressible MHD.
Also the viscous force is now more complex.
Hence, we have
\EQA
\label{dAA}
\DDD^{A}\AAA&=&\UU\times\BB+\FFM+\eta\nab^2\AAA,\\ 
\!\!\!\!\!\rho(\DDD^{U}\! +\UU\!\cdot\!\nab)\UU+\nab P&=&\JJ\times\BB+\rho \FFK+ \nab\!\cdot\!(2\nu\rho\boldsymbol{\mathsf S}), \\ 
(\DDD+\UU\cdot\nab)\ln\rho&=&-\nab\cdot\UU,\label{dlr2}
\ENA
where $\JJ=\nab\times\BB$ is the current density
with the vacuum permeability set to unity.
Furthermore,
\EQA
\DDD^{A} \AAA&=&\DDD\AAA+S A_y \xxx, \\
\DDD^U \UU&=& \DDD\UU +S U_x \yyy + 2 \OO\times\UU, \\
\DDD &=& \partial_t + S x \partial_y
\ENA
are linear operators.
Here,  
$\OO=\Omega \zzz$
is the global angular velocity vector, 
${\sf S}_{ij}=(U_{i,j}+U_{j,i})/2-\delta_{ij}\nab\cdot\UU/3$   
are the
components of the traceless
rate-of-strain tensor $\SSSS$, where commas denote
partial differentiation, 
and $P$ is the pressure related to the density
via $P=\cs^2\rho$ with 
the isothermal sound speed $\cs$.
Terms containing $S$ are due to a linear background shear flow of the form $\UUS=Sx\yyy$.
Magnetic diffusivity $\eta$, kinematic viscosity $\nu$, and sound speed are assumed constant.
For $\cs^2\ln\rho$ (``pseudo enthalpy") we shall employ the symbol $H$.

Throughout, we define mean quantities by horizontal averaging, i.e., averaging over $x$ and $y$, denoted by an overbar. 
So the means
depend on $z$ and $t$ only.
Fluctuations are denoted by lowercase symbols or primes, e.g.,
$\aaaa=\AAA-\meanAA$, 
$(\uu\times\bb)'=\uu\times\bb-\overline{\uu\times\bb}$,
and $\bm{f}_{\rm K,M}=\bm{F}_{\rm K,M}-\overline{\bm F}_{\rm K,M}$.
The horizontal average obeys the Reynolds rules,
given that $\UUS$ can effectively be treated as a mean flow. For peculiarities involved here, we refer to \cite{SMHD}, Sec.\,2.3.1.

\blue{\subsection{Compressible test-field method (CTFM)}}

The 
\blue{starting point for establishing any TFM are the}
evolution equations for the fluctuating quantities, here of 
magnetic vector potential, 
$\aaaa$,  
velocity, $\uu$, and 
pseudo-enthalpy, $h =\cs^2 (\ln\rho)'$, which
follow from \eqss{dAA}{dlr2} as
\EQA
\DDD^{A}\aaaa&=&  \meanUU\times\bb+\uu\times\meanBB+(\uu\times\bb)'+\ffM+\eta\nab^2\aaaa,\label{dafl}\\
\DDD^{U}\uu  &=&- \nab h + \rho_{\rm ref}^{-1} \left[\,\meanJJ\times\bb+\jj\times\meanBB+(\jj\times\bb)' \right] \nonumber\\ && 
 - \meanUU\cdot\nab\uu -  \uu\cdot\nab\meanUU -(\uu\cdot\nab \uu)' + \ffK  \label{dufl} \\
 &&+ \nu\!\left(\nab^2\uu +\nab \nab\cdot\uu/3\right) \nonumber\\
  &&+ 2\nu\big[ \meanSSSS \cdot\nab h  + \ssss \cdot\nab \meanH + (\ssss \cdot\nab h)' \big]/\cs^2 \nonumber \\
 \DDD h &=& - \meanUU\cdot \nab h - \uu\cdot\nab \meanH  - (\uu\cdot\nab h)'- \cs^2\nab\cdot\uu.\label{dhfl}
\ENA
In order to avoid the occurrence of triple correlations, we have, however,
modified the momentum equation by replacing the density in the denominator
of the Lorentz acceleration by a reference density $\rho_{\rm ref}$.
It is set equal to the volume averaged density and is
constant in time as mass is conserved.
A possible refinement would consist in using a horizontal 
average instead, thus allowing $\rho_{\rm ref}$ to change in time and to depend on $z$.

\blue{
\subsubsection{The zero problem}
}

In the QKTFM (see Sect.~\ref{sec:qktfm}), 
the mean electromotive force 
$\meanEMF = \overline{\bm{u} \times \bm{b}}$,
is a functional of only $\uu$, $\meanUU$, and $\meanBB$ (linear in $\meanBB$).  
However, in the more general case with a magnetic background turbulence,
this is no longer so.
To deal with this difficulty, RB10
added the  evolution equations for the background turbulence
$\left(\uuz,\bbz\right)$ to the equations of the TFM.
In the context of 
\blue{
(isothermal)
}
FMHD
we add the evolution equations for
$\left(\uuz,\bbz,\hz\right)$,
which by definition apply for 
zero mean field
--- thus we name this system 
the ``zero problem".
Let now all variables be split
in parts independent of (superscript ``$(0)$") and 
\blue{vanishing with}
$\meanBB$ (superscript ``$(B)$"), respectively,
like $\uu=\uuz+\uuB$ etc.
Then
\onecolumngrid
\EQA
\DDD^{A}\aaB&=&  \meanUU\times\bbB+\uu\times\meanBB+ \left(\uuz\times\bbB + \uuB\times\bbz + \uuB\times\bbB \right)'+\eta\nab^2\aaB,\label{daT}\\
\DDD^{U}\uuB  &=&- \nab \hB + \rho_{\rm ref}^{-1} \left[\meanJJ\times\bb+\jj\times\meanBB+\left(\jjz\times\bbB + \jjB\times\bbz + \jjB\times\bbB\right)' \right]\nonumber\\ && 
 - \meanUU\cdot\nab\uuB -  \uuB\cdot\nab\meanUU - \left(\uuz\cdot\nab \uuB + \uuB\cdot\nab \uuz + \uuB\cdot\nab \uuB \right)'   \label{duT} \\
 &&+ \nu\!\left(\nab^2\uuB +\nab \nab\cdot\uuB/3\right) + 2\nu\left[ \meanSSSS \cdot\nab \hB  + \ssB \cdot\nab \meanH + \left(\ssz \cdot\nab \hB + \ssB \cdot\nab \hz + \ssB \cdot\nab \hB \right)' \right]/\cs^2 \nonumber \\
 \DDD \hB &=& - \meanUU\cdot \nab \hB - \uuB\cdot\nab \meanH  - \left(\uuz\cdot\nab \hB + \uuB\cdot\nab \hz + \uuB\cdot\nab \hB\right)'- \cs^2\nab\cdot\uuB ,\label{dhT}
\ENA
and the ``zero problem" is given by
\EQA
\DDD^{A}\aaz&=&  \meanUU\times\bbz+\left(\uuz\times\bbz\right)'+\ffM+\eta\nab^2\aaz,\label{daTfl}\\
\DDD^{U}\uuz  &=&- \nab \hz + \rho_{\rm ref}^{-1} \left(\jjz\times\bbz \right)' \nonumber\\ && 
 - \meanUU\cdot\nab\uuz -  \uuz\cdot\nab\meanUU - \left(\uuz\cdot\nab \uuz \right)' + \ffK  \label{dufl} \\
  &&+ \nu\!\left(\nab^2\uuz +\nab \nab\cdot\uuz/3\right) + 2\nu\left[ \meanSSSS \cdot\nab \hz  + \ssz \cdot\nab \meanH + \left(\ssz \cdot\nab \hz\right)' \right]/\cs^2 \nonumber \\
 \DDD \hz &=& - \meanUU\cdot \nab \hz - \uuz\cdot\nab \meanH  - \left(\uuz\cdot\nab \hz\right)'- \cs^2\nab\cdot\uuz \label{dhfl}.
\ENA
\twocolumngrid
Note that, while \Eqs{daT}{duT} are   
visibly inhomogeneous\footnote{The more precise term is ``heteronomous" as used in the nonlinear dynamical systems context.}
via the terms $\uuz\times\meanBB$ and $\meanJJ\times\bbz+\jjz\times\meanBB$ (both homogeneous in $\meanBB$), this also holds true
for \Eq{dhT} via $\nab\cdot\uuB$, which does not vanish for $\meanBB\ne\bm{0}$.

In general, $\meanEMF=\overline{\uu\times\bb}$ can be split  into a contribution
$\meanEMF^{(0)}=\overline{\uuz\times\bbz}$, which is independent of the mean field,
and
\EQ
\meanEMF^{(B)}=\overline{\uuz\times\bbB}
+\overline{\uuB\times\bbz}
+\overline{\uuB\times\bbB},  \label{emfB}
\EN
where $\uuB$ and $\bbB$ denote the solutions of \eqs{daT}{duT}.
\blue{
In the presence of an SSD, 
$\meanEMF^{(0)}$ is commonly 
expected to vanish, hence we don't consider it any further.}

\blue{
\subsubsection{The kinematic limit}
In the kinematic limit, the mean 
magnetic field evolving  in the main run is too weak to cause any significant deviations of the fluctuating fields 
from the background turbulence, that is, $\uu \rightarrow \uuz$, $\bb \rightarrow \bbz$,
and $h  \rightarrow \hz$.
Correspondingly, to obtain all unknowns as first-order quantities in $\meanBB$,
terms like $\big(\uuB\times\bbB\big)'$ in \Eqss{daT}{dhT} need to be dropped, 
and in $\uu\times\meanBB$, $\jj\times\meanBB$, $\meanJJ\times\bb$, the fluctuations $\uu,\jj,\bb$ need to be replaced by their counterparts from the ``zero problem",
$\uuz,\jjz,\bbz$. Thus, we obtain
\onecolumngrid
\EQA
\DDD^{A}\aaB&=&  \meanUU\times\bbB+\uuz\times\meanBB+ \left(\uuz\times\bbB + \uuB\times\bbz \right)'+\eta\nab^2\aaB,\label{daTk}\\
\DDD^{U}\uuB  &=&- \nab \hB + \rho_{\rm ref}^{-1} \left[\meanJJ\times\bbz+\jjz\times\meanBB+\left(\jjz\times\bbB + \jjB\times\bbz \right)' \right]\nonumber\\ && 
 - \meanUU\cdot\nab\uuB -  \uuB\cdot\nab\meanUU - \left(\uuz\cdot\nab \uuB + \uuB\cdot\nab \uuz \right)'   \label{duTk} \\
 &&+ \nu\!\left(\nab^2\uuB +\nab \nab\cdot\uuB/3\right) + 2\nu\left[ \meanSSSS \cdot\nab \hB  + \ssB \cdot\nab \meanH + \left(\ssz \cdot\nab \hB + \ssB \cdot\nab \hz \right)' \right]/\cs^2 \nonumber \\
 \DDD \hB &=& - \meanUU\cdot \nab \hB - \uuB\cdot\nab \meanH  - \left(\uuz\cdot\nab \hB + \uuB\cdot\nab \hz \right)'- \cs^2\nab\cdot\uuB ,\label{dhTk}
\ENA
\twocolumngrid
This system is an inhomogeneous linear system for the variables $\aaB,\uuB,\hB$, with its inhomogeneities being in turn linear and homogeneous in the mean field $\meanBB$.
Hence, disregarding the influence of initial conditions, there are solutions being linear in $\meanBB$ and vanishing for $\meanBB=\boldsymbol{0}$.
This qualifies \Eqss{daTk}{dhTk} directly for being cast into a test-field procedure.
As a caveat, we should mention that this system may have non-vanishing solutions for $\meanBB=\boldsymbol{0}$, namely unstable eigenmodes of its homogeneous part.
}

\blue{
In the kinematic limit, the mean EMF $\meanEMF^{(B)}$, the mean ponderomotive force $\meanFFFF^{(B)}$ and the ``magnetically induced mass source" $\meanQ^{(B)}$
have likewise to become linear in the fluctuations $\bbB,\uuB,\hB$, so we write
\begin{align}
\meanEMF^{(B)}&=\overline{\uuz\times\bbB} +\overline{\uuB\times\bbz}, \label{meanEMFBbar}\\
\intertext{for the contribution to $\meanFFFF^{(B)}$ from the Lorentz force}
&\overline{\jjz\times\bbB} +\overline{\jjB\times\bbz}, \label{meanFFFBbar1}
\intertext{(the factor $1/\rho_{\rm ref}$ again omitted here),
for that resulting from self-advection}
&-\overline{\uuz\cdot\nab\uuB} -\overline{\uuB\!\cdot\nab\uuz} \label{meanFFFBbar2}
\intertext{and that resulting from the nonlinear viscous force part}
 &\overline{\ssz \cdot\nab \hB + \ssB \cdot\nab \hz}. \label{meanFFFBbar3}
\end{align}
Finally,  from the advective term in the continuity equation, we obtain
\begin{align}
\meanQ^{(B)} =
&-\overline{\uuz\cdot\nab \hB} - \overline{\uuB\cdot\nab \hz}. \label{meanQBbar}
\end{align}
}

\blue{
\subsubsection{Test fields and parameterizations}
}

We solve \eqss{daT}{dhT} not
by setting $\meanBB$ to the actual mean field resulting from the solutions of
\eqss{dAA}{dlr2}, but by setting it to one out of four test fields
$\BB^{(i)}$, $i=1,\ldots,4$.
Those are
\EQA
&\BB^{(1)}=(\cos \kB z,0,0),\quad \label{tf1}
&\BB^{(2)}=(\sin \kB z,0,0),\\
&\BB^{(3)}=(0,\cos \kB z,0),\quad \label{tf2}
&\BB^{(4)}=(0,\sin \kB z,0),
\ENA
where $\kB$ is the wavenumber of the test field, being a multiple of 
of the wavenumber corresponding to the 
vertical extent
of the computational domain.
From the solutions of \eqss{daT}{dhT} we can construct, 
for each $\BB^{(i)}$ (superscripts $(i)$ left out hereafter),
the mean electromotive forces
$\meanEMF^{(B)}=\big(\,\overline{\uu\times\bb}\,\big)^{(B)}$, the mean ponderomotive forces,
$\meanFFFF^{(B)}=\big(\,\overline{\jj\times\bb/\rho_{\rm ref} - \uu\cdot\nab \uu +  2\nu\,\ssss \cdot\nab h}\,\big)^{(B)}$,
and the mean mass sources
$\meanQ^{(B)}=-\big(\,\overline{\uu\cdot\nab h}\,\big)^{(B)}$
according to \Eqss{meanEMFBbar}{meanQBbar},
and express them in
terms of the mean field by the ansatzes
\EQA
\meanemf_i^{(B)}&=&\alpha_{ij}\meanB_j-\eta_{ij}\meanJ_j,\label{alpeta}\\
\meanFFF_i^{(B)} &=&\phi_{ij}\meanB_j-\psi_{ij}\meanJ_j,\label{phipsi}   \\
\meanQ^{(B)} &=& \sigma_i \meanB_i-\tau_i \meanJ_i,
\ENA
where $i,j$ adopt only the values $1,2$ as a consequence of 
$\meanJ_z=0$ and
setting 
(the anyway constant)
$\meanB_z$ arbitrarily to zero.
Hence, each of the four tensors, $\alpha_{ij}$, $\eta_{ij}$, $\phi_{ij}$,
$\psi_{ij}$, has four components,and 
together with the vectors $\sigma_i$ and $\tau_i$, we have 20 unknowns.
$\alpha_{ij}$, $\phi_{ij}$, and $\sigma_i$ are pseudo-quantities, $\eta_{ij}$, $\psi_{ij}$, and $\tau_i$ true ones.
Note that often the $\alpha$ and $\eta$ tensors are defined as just the symmetric parts of our $\alpha_{ij}$ and $\eta_{ij}$
while their antisymmetric parts are cast into the vectorial coefficients of the $\gamma$ and $\delta$ effects. 
The coefficients $\boldsymbol{\alpha}$, $\boldsymbol{\beta}$, $\boldsymbol{\gamma}$, and $\boldsymbol{\delta}$
describe in turn the effects of turbulent generation, diffusion, pumping and the (non-generative, non-dissipative) so-called R\"adler effect.
In the presence of shear, the coefficient $\eta_{yx}$ plays a prominent role; see \Sec{shear}.

In spite of what could be expected from the Lorentz force, being quadratic in $\BB$, the turbulent ponderomotive force \eq{phipsi}
is with leading order
linear in $\meanBB$. This is because of the 
effect
of the magnetic background turbulence $\bbz$ via, in the kinematic limit, $\overline{\jjz\times\bbB + \jjB\times\bbz}$.
A mean mass source $\meanQ^{(B)}$ due to non-vanishing vectors $\ssigma$ or $\ttau$ requires anisotropy of the turbulence. It was not 
considered in this work,
nor was $\meanFFFF^{(B)}$
\blue{
due to non-vanishing $\pphi$ and $\ppsi$.
}

\blue{
\subsubsection{The nonlinear case}
\label{sec:nonlinear}
}

\blue{
In the QKTFM, only the 
velocity matters for the turbulent transport coefficients and can readily be identified with the one of the ``main run",  
i.e. the system \eqref{dAA}--\eqref{dlr2} solved simultaneously
with the test problems.
Thus, the opportunity opens to deal with quenching, that is the effect of the evolving mean field in the main run onto the fluctuating velocity and thus 
the coefficients. 
A try to proceed analogously in the CTFM encounters a threefold difficulty:
First,  \Eqss{daT}{dhT} are in general nonlinear PDEs, thus conflicting with the requirement that the  coefficients measured by a TFM have to be independent of the
amplitude of the test field, which implies linearity.
Second, even when dropping the terms quadratic and bilinear in $\uuB, \bbB, \hB$, these variables would show nonlinear dependences on the amplitude of $\meanBB$
by virtue of terms of the form $\uuB\times \meanBB$ etc.
Third, there is no obvious channel through which the fluctuating quantities of the main run
which carry  the imprint of the evolving mean field of the LSD,
 would enter the system \eqref{daT}--\eqref{dhT}.}

\blue{
All three difficulties can be overcome by a trick, which, however, compromises mathematical rigor:
by identifying $\uu,\bb,h$ {\it partly} with the corresponding quantities $\uuM,\bbM,\hM$ {\it of
the main run} in such a way, that the system \eqref{daT}--\eqref{dhT}
becomes formally linear and its solutions linear functionals of $\meanBB$.
Rigor is lost as in general $\uuM\ne\uuz+\uuB$ etc.
While it is inevitable to replace $\uu$ by $\uuM$ in $\uu \times \meanBB$, $\bb$ by $\bbM$ in $\meanJJ\times\bb$ and $\jj$ by $\jjM$ in $\jj\times\meanBB$, 
there are several choices to reform the bilinear and quadratic terms
${(\uu\times\bb)'}^{(B)}$, ${(\jj\times\bb)'}^{(B)}$, ${(\uu\cdot\nab \uu)'}^{(B)}$, ${(\ssss \cdot\nab h)'}^{(B)}$, and ${(\uu\cdot\nab h)'}^{(B)}$.
For example, ${(\uu\times\bb)'}^{(B)}$ can be rewritten as
\begin{align}
&\big(\uuM\times\bbB\big)'
+\big(\uuB\times\bbz\big)' \quad\text{or} \label{EEEBdash} \\
&\big(\uuz\times\bbB\big)'
+\big(\uuB\times\bbM\big)' \,.
\nonumber
\end{align}
Likewise, one writes the fluctuating Lorentz force as
\begin{align}
&\big(\jjM\times\bbB\big)'
+\big(\jjB\times\bbz\big)' 
\quad\text{or}\label{FFFBdash1}\\
&\big(\jjz\times\bbB\big)'
+\big(\jjB\times\bbM\big)',
\nonumber
\end{align}
(the factor $1/\rho_{\rm ref}$ omitted here).
The fluctuating self-advection term and
the fluctuating nonlinear viscous force part,
as well as the fluctuating part of the advective term in the continuity equation
are rewritten in an analogous way.
All these versions are linear in quantities with superscript $\meanBB$.
Identifying $\meanBB$ with any one out of a set of test fields, we call the systems \eqref{daTk}--\eqref{dhTk} and \eqref{daT}--\eqref{dhT} with any combination of the above rearrangements applied, the test problems and their solutions the test solutions.
As we have five bilinear/quadratic terms, altogether 32 different versions (``flavors") of the CTFM exist,
out of which, however, we consider only the four, already employed in earlier works.
Note that the different flavors have in general 
different stability properties.}
\blue{
For most of the runs of this work, we chose to use
the respective first versions of \Eqs{EEEBdash}{FFFBdash1} etc.
This choice corresponds to what is called the {\sf ju} flavor; see Table~1 of RB10.
In some tests we also employ the second versions of \Eqs{EEEBdash}{FFFBdash1} (the {\sf bb} flavor).}

\blue{
\subsubsection{Construction of functionals linear in $\overline{\boldsymbol{B}}$ }
}

\blue{In the kinematic limit $\meanBB\rightarrow \bm{0}$,
 all flavors of the CTFM 
 converge and
have thus to yield identical results up to roundoff errors.}

\blue{
To guarantee that the mean EMF $\meanEMF^{(B)}$, the mean ponderomotive force $\meanFFFF^{(B)}$,
and the mean ``magnetically induced mass source" $\meanQ^{(B)}$
become also formally linear functionals of $\meanBB$ one proceeds 
analogously to the treatment of the fluctuating parts of the bilinear/quadratic terms, e.g.
\begin{align}
\meanEMF^{(B)}&=\overline{\uuM\times\bbB}
+\overline{\uuB\times\bbz} \quad \text{or} \label{meanEMFBbar}\\
&=\overline{\uuz\times\bbB}
+\overline{\uuB\times\bbM},
\nonumber
\end{align}
cf. \ Eqs.~(29) and (30) of RB10.
Henceforth we drop the superscript `mr' so that quantities without superscript always refer to the main run.}

\blue{
\subsubsection{Remarks}
\begin{enumerate}
\item The major advance afforded by the new method is to provide a tool, which is reliable when magnetic
background turbulence is present 
and compressibility is fully taken into account,
albeit restricted to isothermality.
However, it deals with nonlinearity in the same way as the earlier methods for this class of problems. 
\item
An analogous, yet simpler TFM can be established for incompressible MHD, and the theoretical basis for that was laid out in
Appendix~A of RB10.
\item
Higher than second-order correlations are already entering the transport
coefficients in the kinematic limit \big(then of the fluctuations $\bbz$,
$\uuz$, $\hz$\big) beyond SOCA, and it is one of the strengths of any
TFM not to be restricted to a certain maximal correlation order.
\item
Given the lack of mathematical rigor of the nonlinear version of the method,
agreement of all 
or at least some of
its different flavors
may provide a heuristic argument for correctness.
\item
The transport coefficients delivered by the nonlinear version are dependent on the mean field
{\em in the main run}, but not on the amplitude of the test fields.
Yet, for the general case of poor scale separation, the coefficients do depend on the {\it scale} of the test fields.
Hence, for 
a meaningful
application to the interpretation of the main run, it is important
to guarantee that the dominant
scale of the mean field observed in it agrees with that of the test fields.
That granted, the coefficients can be employed to establish a mean-field model of the main run, the validity of which, however,
is limited to just the observed mean field; see \cite{TB08} for an
example illustrating such a limitation.
Predicting correctly a zero growth rate from such a model for a main run with saturated
mean field on an MHD background would provide a strong argument for the correctness
of the nonlinear version; see \cite{BRRS08} for such a study regarding the QKTFM,
We refer this to future work.
\end{enumerate}
}

\blue{
\subsubsection{Mean-field equations}
}

For completeness, we provide here the equations governing the mean quantities $\meanAA$, $\meanUU$ and $\meanH$
\begin{align}
  \partial_t \meanAA =  &-S \meanA_y \xxx + \meanUU\times\meanBB + \eta \nab^2 \meanAA + \meanEEEE, \\
  \partial_t \meanUU = &-S \meanU_x \yyy -\nab \meanH - \meanUU\cdot\nab \meanUU - 2\OO\times\meanUU \\
  & +  \nu (\nab^2 \meanUU + \nab\nab\cdot\meanUU/3) + 2\nu \meanSSSS \cdot\nab \meanH/\cs^2  \\
  & + \meanJJ\times\meanBB/\rho_{\rm ref} + \meanFFFF,\\
  \partial_t \meanH   = &- \meanUU\cdot\nab \meanH - \cs^2\nab\cdot \meanUU + \meanQ\,.
\end{align}
Note that in this most general form, all quantities comprise `$(0)$' and `$(B)$' constituents. Thus, also the vorticity
dynamo is covered, which to model, however, one would need a parameterization of  $\meanFFFF^{(0)}$ and $\meanQ^{(0)}$ in terms of
$\meanUU^{(0)}$. The CTFM cannot produce such.

\subsection{Quasikinematic TFM}\label{sec:qktfm}

We state here for comparison the governing equation for the QKTFM
\cite[see also][]{Schrinner05,Schrinner07},
which is just \eq{daT} with $\bbz=\bm{0}$,
while dropping \eq{duT} -- \eq{dhT}.
Hence \eq{meanEMFBbar} reduces simply to
\EQ
\meanEMF^{(B)}=\overline{\uu\times\bbB}
\EN
and we find the contribution $\overline{\uuB\times\bbz}$ missing.
Again, for further details see RB10.

\subsection{Resetting}
\label{Resetting}

The test problems \eqss{daT}{dhT},
being linear,
can
have unlimitedly growing solutions,
but usually the measured transport coefficients
nevertheless show intervals, in which they are
statistically stationary,
in other words show ``plateaus".
If these are absent altogether,
we disregard such a measurement, 
or try to improve it by lowering the time step or increasing the resolution.
We reset the test 
solutions
after regular intervals (typically every 15--20
turnover times in this study); see \cite{Hub+09} for a discussion.
Each resetting interval hence contains an initial transient, which we
remove from the analysis.
If the coefficients also show non-stationary behavior 
towards the end of the resetting interval, these parts are 
removed, too.

\blue{
\subsection{Comments on fluctuations and averages}
}

\blue{
The CTFM yields coefficients,
which still depend on $z$ and $t$.
We usually present them as quantities that are additionally averaged
over
these coordinates,
but we emphasize that the fluctuations in $z$ and $t$
can be 
themselves
an intrinsic component of another class of dynamos, for example the
incoherent $\alpha$--shear dynamos \citep{VB97}.
The fact that the mean-field coefficients, and 
hence
also the mean fields,
fluctuate in those remaining coordinates is a
common feature of mean-field theory in cases
lacking
strong separation between
the spatio-temporal scales of mean and fluctuating quantities.
Dealing correctly with limited scale separation requires 
to take nonlocal effects properly into account \citep{BRS08,RB2012,BC18},
but this 
does not 
at all
compromise the usefulness of mean-field theory.
The CTFM is without any modifications capable of dealing with scale dependence
w.r.t.\ $z$ via varying the wavenumber $\kB$,
and in this paper we apply the kinematic CTFM to study scale dependence
in the shear dynamo
context.
For temporal nonlocality, merely time-dependence (harmonic or exponential)
has to be added to the specification of the test fields \eqref{tf1},\eqref{tf2}.
}

\subsection{Random forcing} \label{forcing}

We utilize the standard random forcing as implemented in 
the {\sc Pencil Code} \citep{PC2020},
which employs white-in-time harmonics.
Their wavevectors are chosen from a thin shell in $\kk$ space of radius $\kf$ 
further requiring that they fit into the computational domain.
We also exclude the case $k_y=0$ to avoid a  
mean field or flow to be  directly sustained.

\subsection{Suppression of the vorticity dynamo} \label{vortdyn}

The runs with shear alone ($\Omega=0$)
are prone to a hydrodynamic
instability, leading to the generation of 
mean flows and usually referred to as the vorticity dynamo \citep[see, e.g.][]{Elp03,KMB09}. As in 
\cite{SMHD},
we prefer to suppress these flows, to focus on studying the magnetic shear 
current effect in isolation. The procedure adopted 
there, namely subtracting 
$xy$
averaged mean flows,  
is not a sufficient measure here
 as not all the mean flows are captured.
Hence, we turn to
another method, namely suppressing the vorticity dynamo by adding a small amount
of rotation to the system. For $\Omega$ and $S$ of opposite sign, which 
is the standard case in 
galactic and accretion disks,
we would be limited to the range where
$q\equiv-S/\Omega<2$, 
 as at the upper limit the flow
would become Rayleigh unstable. If we, however, choose the same sign, hence 
a negative 
$q$, 
we can avoid this limitation.
Here, we investigate values of $q$ in the range $[-40,...,-10]$, and
choose the maximum value that still suppresses the vorticity dynamo, but does not yet
significantly affect the test field measurements; see \Sec{hydro}.

\subsection{Input and output quantities}

The simulations are fully defined by choosing shear parameter $S$, rotation rate $\Omega$,
the forcing amplitude and wavenumber, $\kf$, and the diffusivities $\nu$ and $\eta$.
For normalizations we use the length scale $k_1^{-1}$, with $k_1=2\pi/L$,
where $L$ is the extent of the simulation domain in any direction,
and the acoustic time scale $\tau_s=(c_s k_1)^{-1}$.
Rotation rate $\Omega$ and shear
 rate
 $S$ 
are normalized by the acoustic time scale as
$\Ot=\Omega \tau_s$ and $\Sti=S \tau_s$.
The boundary conditions are 
periodic in $y$ and $z$, while shearing--periodic in $x$.

For the velocity field, we define a time-averaged root-mean-square (rms) value as
$\urms=\big\langle \langle {\bm u}^2\rangle^{1/2}\big\rangle_{t}$, and a time-dependent variant
$\urmst= \langle {\bm u}^2\rangle^{1/2}$.  
Simulation results are often shown as functions of the time in units of
turnover time, $t \urms \kf$.  
Similarly, we define rms values for the magnetic field
$\Brms= \big\langle \langle {\bm B}^2\rangle^{1/2}\big\rangle_{t}$,
and
$\Brmst = \langle {\bm B}^2\rangle^{1/2}$, 
while
$\meanB_{i,{\rm rms}}=\big\langle \meanB_i^2 \big\rangle_z^{1/2}$ are the
rms values of the mean field components. Here,
$\langle.\rangle$ denotes volume averaging and 
$\langle.\rangle_\xi$ 
averaging over a coordinate $\xi$.
The magnetic field is normalized
by the equipartition field strength,
$B_{\rm eq}=\big\langle \langle \rho\uu^2 \rangle ^{1/2}\big\rangle_t$.

For diagnostics, we quantify the strength of the turbulence 
by the fluid and magnetic Reynolds numbers
\begin{equation}
\Rey=\frac{\urms}{\nu \kf},\ \ \ \Rm=\frac{\urms}{\eta \kf}=\Pm \, \Rey,
\end{equation}
where 
\begin{eqnarray}
\Pm=\frac{\nu}{\eta},
\end{eqnarray}
is the magnetic Prandtl number.
The Lundquist number 
is
given by
\begin{equation}
\Lu=\frac{\Brms}{\eta\kf\sqrt{\rho_{\rm ref}}}. 
\end{equation}
The strength of
the imposed shear is measured by the dynamic shear number
\begin{eqnarray}
\ShK=\frac{S}{\urms \kf}.
\end{eqnarray}
We normalize the turbulent magnetic diffusivity tensor by
the molecular diffusivity $\eta$.

\subsection{Interpretation of the dynamo instability}

As in \cite{SMHD}, we compute three different dynamo numbers describing a 
1D mean--field dynamo model\footnote{In \cite{SMHD} we labelled this model ``0D" because the explicit $z$ dependence can be eliminated by employing the ansatz $\meanAA\sim\exp(\ii k z)$},
in which
 both the coherent SC effect and the incoherent ones due to $\alpha_{yy}$ and $\eta_{yx}$ fluctuations 
 with zero mean
 are taken into account. 
The coherent shear current effect is characterized by
\begin{equation}
D_{\eta S} \equiv \frac{1}{\etaT^2} \left[ \left(\frac{S}{k_z^2} + \eta_{xy}\right) \eta_{yx}+\epsilon 
\right], \label{Eq:SC}
\end{equation}
where $\etaT = \eta + \etat$, $\etat = \left( \eta_{xx} + \eta_{yy} \right)\!/2$, $\epsilon =\left( \eta_{xx}-\eta_{yy} \right)\!/2$, and $k_z$ is the 
dynamo wavenumber. 
Our standard approach is to identify this wavenumber from
the Fourier mode growing fastest during the exponential phase of an LSD and
we denote it as $k_{z,{\rm kin}}$.
The incoherent $\alpha$--shear-driven dynamo is described by
\begin{equation}
D_{\alpha S} = \frac{\alpha_{\rm rms} \left| S \right| }{\etaT^2 k_z^3},
\end{equation}
where we consider in $\alpha_{\rm rms}$ only the fluctuations of  $\alpha_{yy}$. 
Finally, for a dynamo driven by the incoherent SC
effect due to fluctuations of $\eta_{yx}$
and shear,
\begin{equation}
D_{\eta_{\rm rms} S} = \frac{\eta_{yx,\rm rms} \left| S \right| }{\etaT^2 k_z^2} \label{Eq:ISC}
\end{equation}
is relevant.
In \cite{SMHD}, we derived the marginal dynamo numbers for a grid of combinations of
$(D_{\eta S},D_{\alpha S},D_{\eta_{\rm rms} S})$, and we refer the reader to these results. 
For orientation, we note that 
in the absence of the incoherent effects,
$D_{\eta S}>1$ is required for dynamo action, but in the
absence of the coherent and incoherent SC effects, $D_{\alpha S} > 2.3$.
The presence of the SC effects can increase $D_{\alpha S}$,
but this influence is mild.

\section{Results}
\enlargethispage{2\baselineskip}

\subsection{Roberts flow and field} \label{roberts}

\begin{figure}[t!]\begin{center}
\includegraphics[width=\columnwidth]{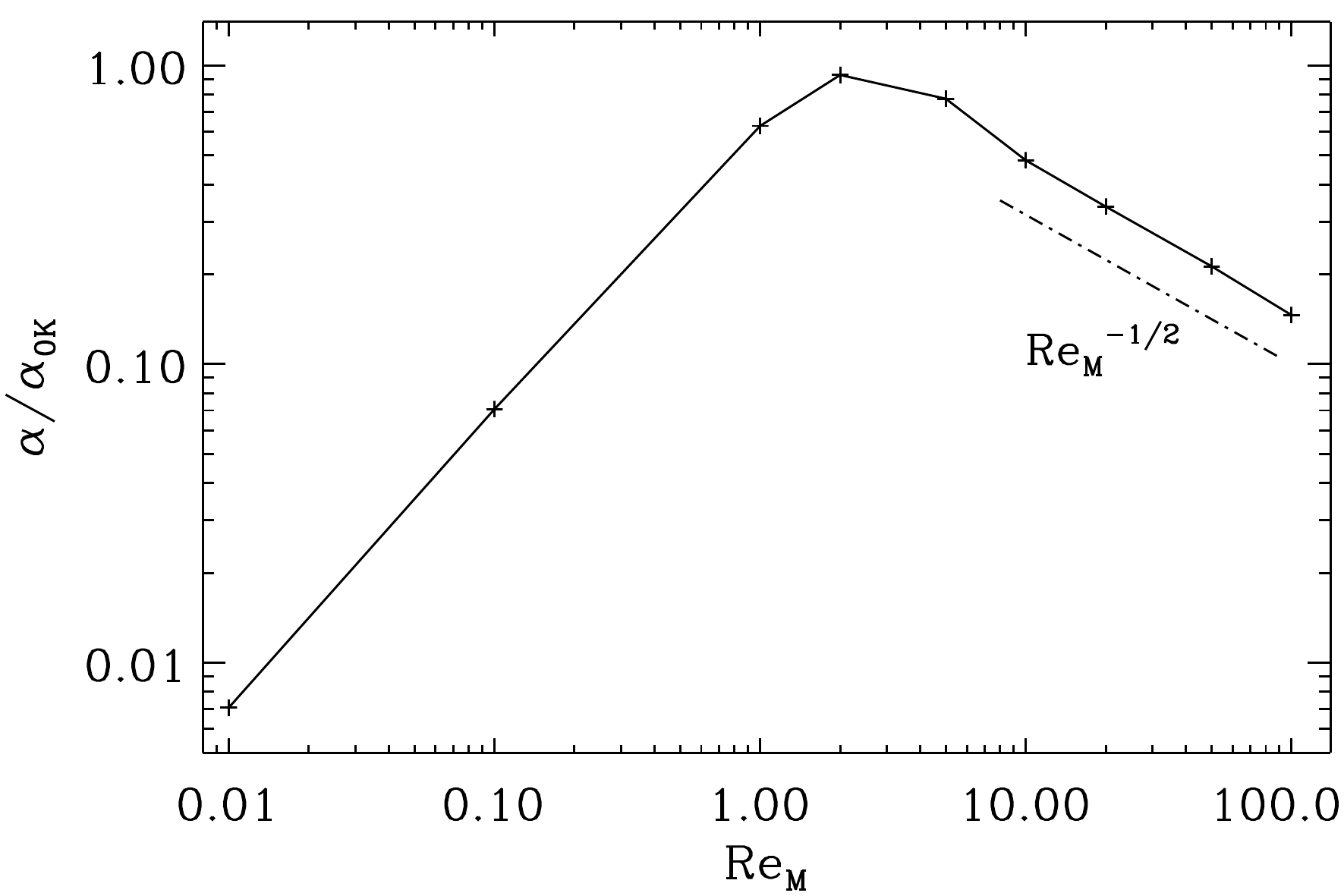}
\end{center}\caption{
$\Rm$ dependence $\alpha_{xx}=\alpha_{yy}\equiv\alpha$, agreeing
with $\alpha_{\rm imp}$ for the forced Roberts flow.
}\label{p01_0}\end{figure}

A simple and reliable way of validating the CTFM
is to restrict oneself to two dimensions
($x$ and $y$) and compare with the imposed-field method, where 
$\alpha_{xx,yy}=\overline{\uu\times\bb}\cdot\BB_0/B_{0x,0y}^2$
and $\BB_0$ is a uniform field,
imposed in either the $x$ or the $y$ direction.
The two-dimensional case corresponds to $k_z=0$, so that no turbulent
diffusion can act and only the $\alpha$ tensor is considered.
For the flow (or field) geometry, we have chosen case I of \cite{Roberts72},
which is a vector field of the form $(-\cos x \sin y, \sin x \cos y, \sqrt{2} \cos x \cos y)$
having the Beltrami property.
A ponderomotive force is constructed such that without magnetic field exactly that geometry
is obtained,
in the background flow $\uuz$,
that is, the distortion by the $\uuz\cdot\nab \uuz$ term is compensated  
and the pressure gradient vanishes as well as
the nonlinear part of the viscous force. 
In the complementary case of magnetic forcing, an EMF with just the
Roberts geometry is sufficient as due to its Beltrami property and the
linearity of the induction equation the resulting Lorentz force is zero, no flow is driven,
and $\bbz$ has exactly the Roberts geometry $\bbRob$.
In \Figs{p01_0}{p0_01} we show the $\Rm$ dependence
of $\alpha_{xx}=\alpha_{yy}\equiv\alpha$ for 
kinetic and magnetic forcing, respectively ($\alpha_{xy,yx}=0$).
As in \cite{RB10}, we have normalized $\alpha$ by $\alpha_{\rm0K}=-\urms/2$
and $\alpha_{\rm0M}=3\brms/4$, respectively.

\begin{figure}[t!]\begin{center}
\includegraphics[width=\columnwidth]{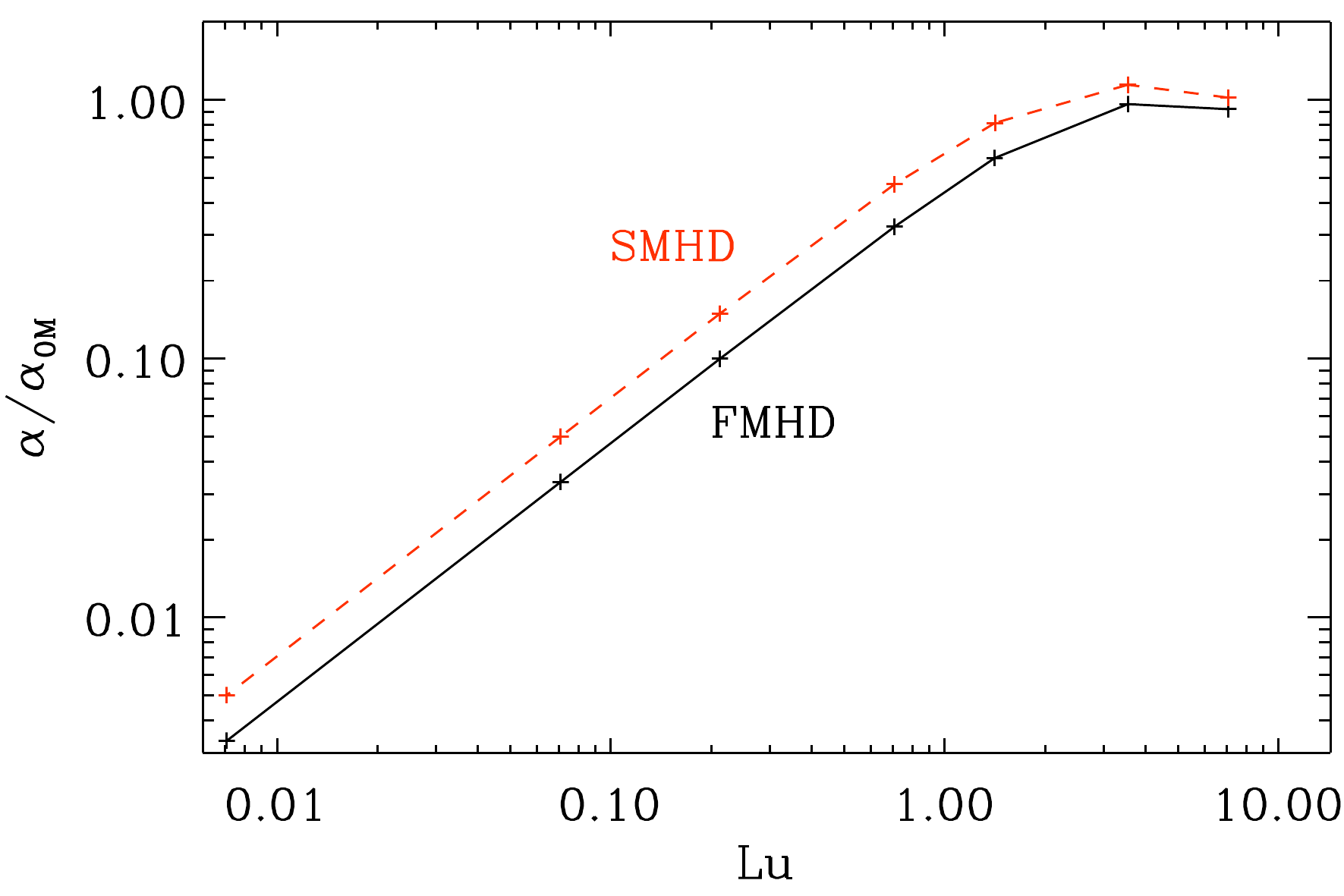}
\end{center}\caption{
$\Lu$ dependence 
of $\alpha$ for forced magnetic background with Roberts geometry
$\Pm=1$,
using FMHD with CTFM and
(black).
Red dashed: SMHD.
}\label{p0_01}\end{figure}

In the kinetically forced case, we compare with the
QKTFM and find perfect agreement, as expected.
In the magnetically forced case, the QKTFM
yields the wrong sign of $\alpha$,
as was already found by \cite{RB10}

in SMHD, while the corresponding TFM was found to
agree with the imposed-field method.
Comparing their results with those of the CTFM,
we find agreement up to some fixed offset for small values of $\Lu$;
see \Fig{p0_01}.

\begin{figure}[t!]\begin{center}
\includegraphics[width=\columnwidth]{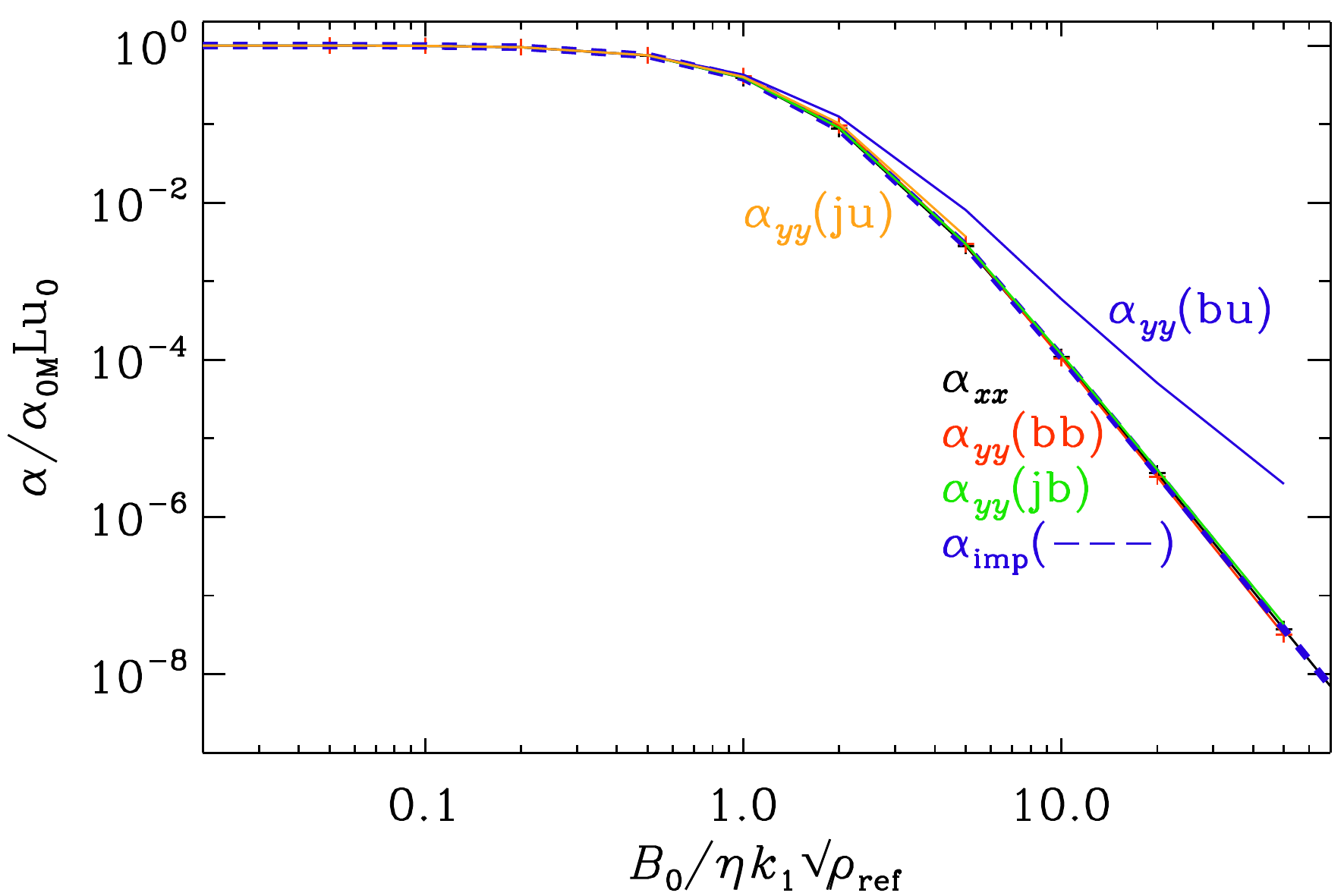} 
\end{center}\caption{
$B_0$ dependence of $\alpha$ for the 
flow-free magnetically forced background with Roberts geometry 
$\big(\bm{0},\bbRob\big)$ and $\Pm=1$ from the CTFM.
\blue{
The results from the imposed-field method are shown by the blue dashed
line and agree with all four flavors of the CTFM for $\alpha_{xx}$.
The other colored solid lines refer to $\alpha_{yy}$ and departures
are seen with the {\sf bu} flavor.
The {\sf ju} flavor (orange) gives even negative values for
$B_0/\bbz_{\rm rms} \gtrsim 7$,
while the {\sf bb} (red) and {\sf jb} (green) flavors
yield $\alpha_{yy}=\alpha_{xx}$.
}
}\label{p0_01_Bdep}\end{figure}

In \Fig{p0_01_Bdep} we show the dependence of $\alpha_{xx}$ and
$\alpha_{yy}$ on $B_0$ for the forced magnetic background with
Roberts geometry, $\nu=\eta$ and forcing amplitudes between 0.01 and 100
\blue{
in units where $\nu=\eta=k_1=1$.
In these cases, flows are only driven by the Lorentz force.
The velocity is generally small compared with $B_0/\sqrt{\rho_{\rm
ref}}$: it can reach 23\% when $B_0/\eta k_1 \sqrt{\rho_{\rm ref}}=1$,
but is smaller both for weaker and stronger fields.
\blue{
Note that this test case, in which the turbulent flow is solely induced by the interaction of the imposed field with the 
magnetic background turbulence $\bbz$, is 
quite different
from the shear dynamo case studied in the later parts of the paper, as well as from
astrophysical settings in general.}
The {\sf jb}, {\sf bb}, {\sf ju}, and {\sf bu} flavors always give
the same results for 
the aligned component
$\alpha_{xx}$, 
also agreeing with those from the imposed field method,
but for strong imposed fields
(in terms of $B_0/b^{(0)}_{\rm rms}$),
the perpendicular component, $\alpha_{yy}$, 
disagrees significantly among the flavors;
see the different lines in \Fig{p0_01_Bdep}.
\blue{We note that the discrepancy 
in $\alpha_{yy}$
is not due to the added compressibility, but was present
already in the simplified MHD case of \cite{RB10}, but was not noticed there.}
As demonstrated in Appendix~\ref{sec:correct}, the correctness in $\alpha_{xx}$ is systematic and extends to arbitrary field strengths.
}
Surprisingly, the slope of the quenching characteristic exhibits now
power $-5$ while in \cite{RB10} and \cite{Karak14} $-4$ had been observed.
It is suggestive to attribute this difference to the inclusion of pressure gradient and self-advection.

\subsection{Shear dynamos with SSD magnetic background turbulence
} \label{shear}

In this section we perform a continuation study of \cite{SMHD}, where 
the SMHD equations were used (governing so-called burgulence)
with kinetic and magnetic forcing.
Now we turn to full MHD,  forced, however, only kinetically
with a nonhelical form of the forcing. 
Hence, in all experiments, we set $\bm{f}_{\rm M}$ to zero. 
We measure the turbulent transport coefficients from volume averaged
quantities, that are, furthermore, averaged over time, neglecting the transients
and untrustworthy parts of the time series, as explained in Sect.\ref{Resetting}. 
The incoherent effects are measured following the same procedure, but rms
values are used: $\alpha_{\rm rms} = \big\langle \langle {\alpha_{yy}}^2\rangle^{1/2}\big\rangle_{t}$
and $\eta_{\rm rms} = \big\langle \langle {\eta_{yx}}^2\rangle^{1/2}\big\rangle_{t}$.
It should be noted that our values of $\alpha_{\rm rms}$ 
and $\eta_{\rm rms}$ underestimate the actual ones, which also include 
the $z$ dependent fluctuations.

\blue{In all the simulations performed in this section, we have used the 
{\sf ju} variant of the CTFM. This choice is based on test runs with the full shear dynamo
setup,
see Appendix~\ref{sheartests}, 
with 
varying magnetic Prandtl numbers ($\Pm=5\ldots20$) 
and $\Sti=-0.2$. These tests revealed that the {\sf ju} and {\sf bb} flavors
gave 
results in good agreement
both in the kinematic and the nonlinear regime with good stability
properties of the test solutions.
The {\sf jb} and {\sf bu} flavors,
however, 
showed 
poorer
stability properties, hence runs employing them
would have required extremely small time steps,
rendering them unfeasible.
These tests indicate that 
the nonlinear CTFM 
\blue{(nCTFM)}
may yield correct results  in
the case of the
shear dynamo
as long as the mean field is
at most slightly above equipartition with the velocity fluctuations.
However,
this is no proof of its general applicability.
}

\subsubsection{Vorticity dynamo}\label{hydro}

\begin{figure*}[h!]\begin{center}
\includegraphics[width=0.8\textwidth]{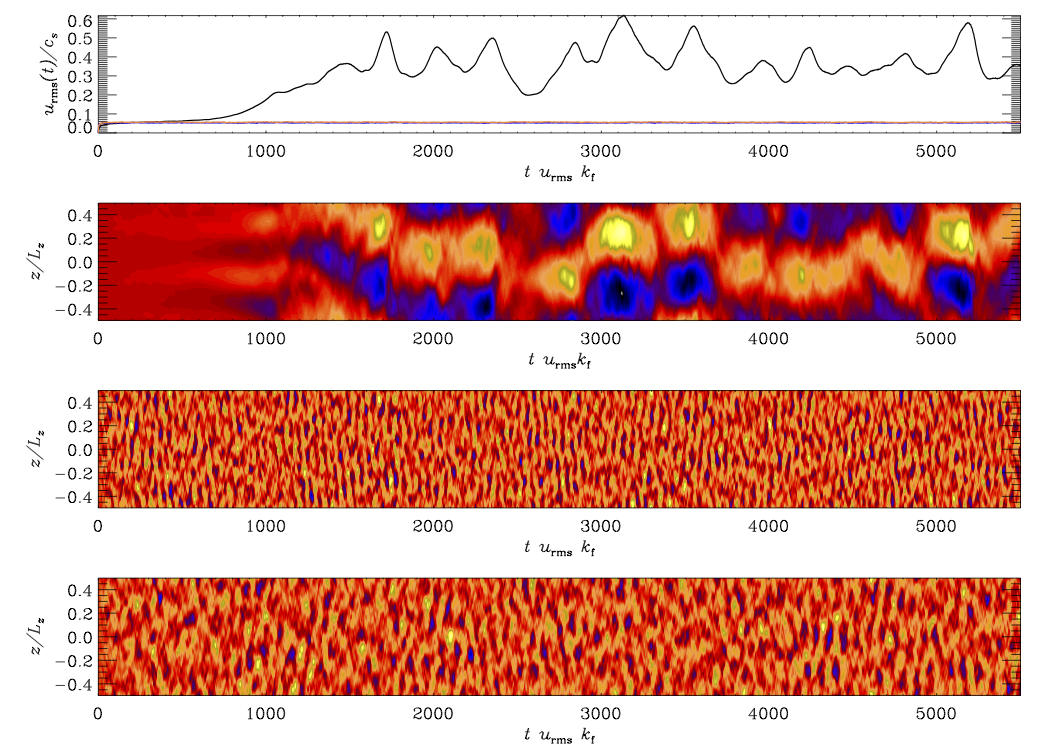}
\end{center}\caption{Comparison of the generated mean flows with $\Sti=-0.2$, $\Ot=0$ (black line panel (a); 
$zt$ diagram in panel (b)), $\Sti=-0.2$, $\Ot=-0.01$ ($q=-20$; blue line in panel (a); (c)), and $\Sti=-0.2$,
$\Ot=-0.005$ ($q=-40$; orange line in panel (a); (d)). The uppermost 
panel shows the time evolution of the volume-averaged rms velocity, $\urmst$. Colors 
in the $zt$ diagrams encode 
$\meanU_y/\cs$ 
with extrema $\pm0.7$,
$\pm0.014$, and $\pm0.022$, in panels (b)--(d), respectively. 
}\label{comprot}\end{figure*}

For certain values of
$\Pm$, namely for 5 and 10, while not for 1 and 20, we see the generation of strong mean flows.
In Figure~\ref{comprot} we 
compare
cases with and without rotation
from purely hydrodynamical counterparts of our MHD runs.
With such experiments we can verify the 
presence of the 
mean flows due to a vorticity dynamo and not due to the  
backreaction of the magnetic field.
In the case of $\Ot=0$, hence shear alone (black line in panel (a), and panel (b)), 
we see the generation of a strong mean flow, which first grows exponentially
with a dominating $k_z=k_1$ mode, and then saturates, exhibiting oscillatory 
behavior with complicated phase migration. 
In the MHD runs with TFM
the mean flows perturb the system to the extent 
that the test solutions start to grow super-exponentially. 
The time step becomes prohibitively small, and no plateaus
can be observed anymore
in the transport coefficients. Hence, all these TFM measurements have been
disregarded.

If a very small amount of rotation is 
added (panels (c) and (d)), the instability 
is suppressed, with $\urmst$ remaining statistically
constant throughout the simulation (blue and orange lines in panel (a)). 
We see, however, that the vertical length scale of the flow is larger in the 
case of weaker 
(d) than in the case of stronger rotation (c). 
This indicates that the $q=-40$ case has still too weak rotation 
to fully suppress the 
vorticity dynamo,
while $q=-20$ 
brings the vertical 
scales close to the forcing scale
$\kf=10 k_1$.
The presence of weak mean flows for $q=-40$
is also reflected in the slightly larger $\urmst$ 
(orange line slightly above the blue one in panel (a)).

\begin{figure}[h!]\begin{center}
\hspace*{-0.8cm}
\includegraphics[width=1.1\columnwidth]{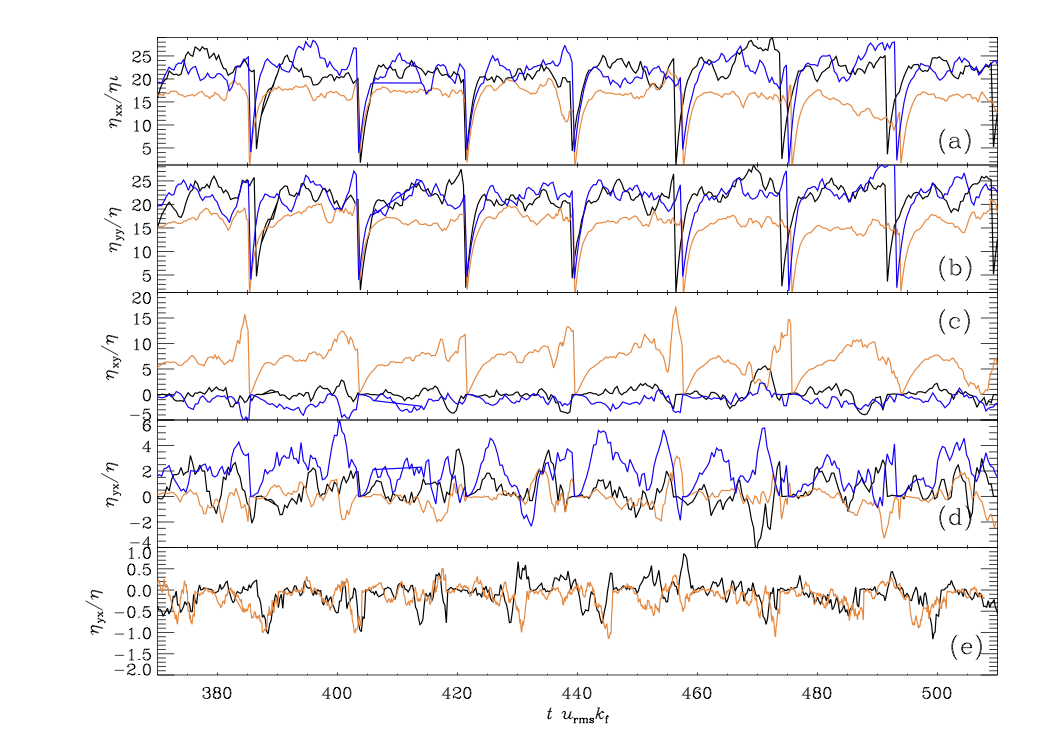}
\end{center}\caption{
Panels (a)--(d):
components of $\eeta$,
measured with the CTFM and normalized to the 
microscopic diffusivity
$\eta$.
Black: 
$\Sti=0$,
$\Ot=-0.01$;     
blue: 
$\Sti=0$,
ten times stronger rotation ($\Ot=-0.1$); 
orange: 
$\Sti=-0.2$, $\Ot=-0.01$  ($q=-20$),
all for $\Pm=10$.
The time axes have been made to match at 
$t\,\urms\kf=400$
hence orange and blue curves 
have been introduced with small offsets
($\Delta t\,\urms\kf=6$ and 8,
respectively).
Panel (e): 
$\eta_{yx}$ from $\Pm=20$ runs with no 
(black) and weak (orange) rotation 
($\Ot=-0.01$, $q=-20$).
}\label{compq}\end{figure}

We have run CTFM
simulations without shear, but with a rotation rate 
corresponding to
$q=-20$, and also with a ten times higher rotation rate; see
Figure~\ref{compq}(a)--(d).
One can observe that the
crucial $\eta_{yx}$ is very similar in the cases with 
weak rotation, with shear included 
(orange lines) or excluded (black lines), while all the other $\eeta$ components are much 
more strongly affected. The diagonal components are clearly larger 
for $\Sti=0$, and $\eta_{xy}$
reverts its sign from strongly positive (with shear) to 
very low values fluctuating about zero (without shear)
\footnote{Note that in the absence of shear but $\Omega\ne0$, the off-diagonal elements of $\eeta$ have to reflect
the R\"adler effect, hence $\eta_{xy}=-\eta_{yx}$. In our results, however, the signal
is drowned out by the fluctuations.
}
From the run with ten times faster rotation (blue lines) we observe
that with increasing rotation rate there is a tendency to
revert the sign of $\eta_{yx}$ to positive and to increase its 
magnitude, while reverting $\eta_{xy}$
to negative values, but with weaker magnitude. We conclude
that for $q=-20$, the rotation rate is still small enough  to not affect the
TFM measurements significantly. Hence,
we select $q=-20$ as the fiducial parameter for our runs.

We use this setup for all Prandtl numbers above unity, to guarantee that no mean flows disturb 
the measurements. We note, however, that for $\Pm=20$ the results are 
nearly identical with and without rotation,
indicating that the vorticity dynamo is not active there.
This is illustrated in Figure~\ref{compq}(e), where we show the measurements of the 
$\eta_{yx}$ component with and without rotation, yielding very similar results in the mid-ranges of the resetting intervals.
The parameters, varied in the henceforth presented simulations,
are the shear rate $\Sti$, the magnetic Prandtl number $\Pm$,
and the forcing wavenumber $\kf$. 

\begin{table*}[h]
\caption{Summary of the kinematic models with variable $\Pm$, shear rate, and vertical wave number of the test fields.}
\hspace*{-2cm}
\begin{tabular}{lrrrrccrrcc} \hline \hline
{Run}  &$\Pm$ &$\Rm$  &$\ShK$ &$\Lu$ & $\eta_{xx}/\eta$ & $\eta_{yy}/\eta$ & $\eta_{yx}/\eta$\phantom{aaa} & $\eta_{xy}/\eta$\phantom{aaa} &$\alpha_{\rm rms}/\eta_t \kf$  &$\eta_{\rm rms}/\eta$\\ \hline \hline
S005Pm1k1 &1 &43 &-0.09 &13 &28.794$\pm$0.322 &28.480$\pm$0.227 &0.122$\pm$0.296 &3.413$\pm$0.413 &0.036$\pm$0.012 &2.601$\pm$0.381 \\
S005Pm1k2 &1 &44 &-0.09 &13 &26.183$\pm$0.317 &25.134$\pm$0.101 &0.468$\pm$1.201 &3.839$\pm$1.100 &0.025$\pm$0.026 &1.282$\pm$1.111\\
S005Pm1k3 &1 &43 &-0.09 &12 &21.763$\pm$0.396 &22.083$\pm$0.313 &-1.707$\pm$1.786 &1.786$\pm$0.556 &0.035$\pm$0.026 &2.241$\pm$2.879\\ \hline 
S005Pm20k1 &20&34 &-0.15 &16 &11.826$\pm$0.095&12.100$\pm$0.048 &0.004$\pm$0.091 &1.127$\pm$0.094 &0.039$\pm$0.020 &0.282$\pm$0.158  \\  
S005Pm20k2 &20&34 &-0.15 &16 &11.754$\pm$0.097&11.982$\pm$0.051 &0.024$\pm$0.080 &1.112$\pm$0.101 &0.039$\pm$0.021 &0.264$\pm$0.182  \\  
S005Pm20k3 &20&34 &-0.15 &16 &11.667$\pm$0.366&11.749$\pm$0.486 &-0.045$\pm$0.043 &1.189$\pm$0.078 &0.038$\pm$0.024 &0.264$\pm$0.189  \\  \hline
S01Pm1k1 &1 &47 &-0.16 &16 &36.436$\pm$0.354 &35.382$\pm$0.098 &0.435$\pm$0.366 &8.707$\pm$0.718 &0.029$\pm$0.015 &2.839$\pm$1.442 \\
S01Pm1k2  &1 &48 &-0.16 &15 &33.434$\pm$0.337 &31.845$\pm$0.348 &0.550$\pm$0.201 &6.879$\pm$0.195 &0.030$\pm$0.018 &2.031$\pm$1.251 \\
S01Pm1k3  &1 &47 &-0.16 &16 &27.735$\pm$0.264 &26.932$\pm$0.335 &0.303$\pm$0.217 &5.271$\pm$0.162 &0.033$\pm$0.019 &1.309$\pm$0.683 \\ \hline 
S01Pm20k1 &20 &32 &-0.30 &27 &7.811$\pm$0.139&8.389$\pm$0.207 &-0.068$\pm$0.066 &0.908$\pm$0.166 &0.040$\pm$0.018 &0.298$\pm$0.283  \\  
S01Pm20k2 &20 &32 &-0.30 &28 &7.678$\pm$0.283&8.151$\pm$0.231 &-0.092$\pm$0.064 &1.063$\pm$0.149 &0.036$\pm$0.020 &0.257$\pm$0.112  \\  
S01Pm20k3 &20 &32 &-0.30 &27 &8.272$\pm$0.255&8.766$\pm$0.350 &-0.002$\pm$0.033 &1.000$\pm$ 0.116 &0.033$\pm$0.018 &0.227$\pm$0.147  \\  \hline \hline 
S02Pm1k1 &1 &66 &-0.23 &25 &70.218$\pm$6.340&65.830$\pm$6.636&0.717$\pm$0.298&32.987$\pm$3.260&0.020$\pm$0.009&4.164$\pm$1.529 \\ 
S02Pm1k2 &1 &66 &-0.23 &25 &56.417$\pm$3.857 &54.361$\pm$3.352&-0.094$\pm$0.742 &22.979$\pm$3.578&0.026$\pm$0.016 &2.663$\pm$0.670\\
S02Pm1k3 &1 &63 &-0.24 &24 &43.668$\pm$2.935 &41.674$\pm$3.695 &-0.132$\pm$0.326 &14.530$\pm$1.093 &0.032$\pm$0.023 &1.478$\pm$0.634\\ \hline 
S02Pm5k1 &5 &16 &-0.48 &17 & 4.149$\pm$1.205 &4.317$\pm$1.542 &0.063$\pm$0.071 &5.192$\pm$0.215 &0.043$\pm$0.017&0.515$\pm$0.143\\
S02Pm5k2 &5 &16 &-0.48 &17 & 4.765$\pm$0.776 &4.576$\pm$0.919 &0.019$\pm$0.070 &4.993$\pm$0.208 &0.036$\pm$0.017&0.370$\pm$0.175\\
S02Pm5k3      &5 &16 &-0.48 &16 &5.444$\pm$0.515 &5.045$\pm$0.578 &0.043$\pm$0.037 &3.180$\pm$0.381 &0.031$\pm$0.013 &0.259$\pm$0.083  \\  \hline
S02Pm10k1 &10 &42 & -0.47 &45 &12.729$\pm$0.989 &12.197$\pm$1.002 &0.059$\pm$0.109 &5.918$\pm$0.179 &0.030$\pm$0.011 &0.787$\pm$0.475 \\ 
S02Pm10k2 &10 &42 & -0.47 &44 &13.203$\pm$0.467 &12.637$\pm$0.611 &0.139$\pm$0.053 &6.440$\pm$0.306 &0.027$\pm$0.011 &0.542$\pm$0.239 \\
S02Pm10k3    &10 &42 &-0.47 &44 &13.416$\pm$0.481&12.514$\pm$0.879 &0.154$\pm$0.039 &5.009$\pm$0.320 &0.027$\pm$0.009 &0.468$\pm$0.165  \\  \hline
S02Pm20k1 &20 &33 &-0.60 &45 &5.438$\pm$0.278 &4.808$\pm$0.138 &0.042$\pm$0.103 &3.282$\pm$0.738 &0.046$\pm$0.019 &0.296$\pm$0.061\\
S02Pm20k2 &20 &33 &-0.60 &45 &5.503$\pm$0.152 &4.900$\pm$0.141 &0.064$\pm$0.084 &4.462$\pm$0.416 &0.045$\pm$0.023 &0.269$\pm$0.083\\
S02Pm20k3 *  &20&33 &-0.60 &44 &6.099$\pm$0.140&5.372$\pm$0.232&0.041$\pm$0.013&1.569$\pm$0.784&0.046$\pm$0.013&0.326$\pm$0.084\\  
S02Pm20k3 *   &20&33 &-0.60 &39 &6.832$\pm$0.077&6.707$\pm$0.392&-0.020$\pm$0.043&2.255$\pm$0.079&0.035$\pm$0.018&0.224$\pm$0.064\\  \hline \hline
S03Pm20k1 &20 &35 &-0.86 &55 &5.711$\pm$0.559 &5.259$\pm$0.627 &-0.043$\pm$0.053 &8.692$\pm$0.458 &0.039$\pm$0.022 &0.240$\pm$0.129\\ 
S03Pm20k2 &20 &35 &-0.86 &56 &5.898$\pm$0.230 &5.242$\pm$0.133 &0.076$\pm$0.072 &7.846$\pm$0.820 &0.038$\pm$0.014 &0.202$\pm$0.153\\ 
S03Pm20k3 &20 &35 &-0.86 &56 &6.274$\pm$0.255 &5.359$\pm$0.161 &0.114$\pm$0.052 &5.511$\pm$0.480 &0.039$\pm$0.022 &0.241$\pm$0.129\\ \hline \hline 
 \end{tabular}\\[2mm]
\label{modelskin}
Note. The run labels are constructed in the following manner: SXXPmYYkZ, where XX indicates the magnitude of the negative $\Sti$, YY the used magnetic Prandtl number $\Pm$, and Z the vertical wave number of the test fields,
$\kB$.
The sets with fixed shear and variable Prandtl number are listed between the two horizontal lines. The forcing wave number is $\kf=5k_1$ for $\Pm$=1 and $\kf=10k_1$ for higher  $\Pm$.
Runs with a star symbol have different SSD solutions.
\end{table*}

\begin{table*}[h]
\hspace*{-1.5cm}
\begin{tabular}{rrrrccrrcc} \hline \hline
Run  &$\Rm$  &$\ShK$ &$\Lu$ & $\eta_{xx}/\eta$ & $\eta_{yy}/\eta$ & $\eta_{yx}/\eta$\phantom{aaa} & $\eta_{xy}/\eta$\phantom{aaa} &$\alpha_{\rm rms}/\eta_t \kf$  &$\eta_{\rm rms}/\eta$\\ \hline \hline
nS00Pm1 &43 & 0 &12 &26.491$\pm$0.321 &26.919$\pm$0.463 &0.094$\pm$0.156 &0.233$\pm$0.429 &0.044$\pm$0.016 &2.873$\pm$0.985\\ 
qS00Pm1 &43 &0 &12 &27.298$\pm$0.207 &26.922$\pm$0.051 &0.110$\pm$0.138 &-0.159$\pm$0.422 &0.045$\pm$0.016 &3.353$\pm$1.423 \\ \hline
{\bf kS005Pm1} &43 &-0.09 &13 &28.794$\pm$0.322 &28.480$\pm$0.227 &0.122$\pm$0.296 &3.413$\pm$0.413 &0.036$\pm$0.012 &2.601$\pm$0.381 \\
{\bf nS005Pm1} &43 &-0.09  &13 &28.516$\pm$0.549 &28.158$\pm$0.502 &-0.151$\pm$0.320 &4.006$\pm$0.125 &0.038$\pm$0.011 &2.810$\pm$0.679 \\
{\bf qS005Pm1} &43 &-0.09 &13 &29.107$\pm$0.377 &28.621$\pm$0.517 &-0.543$\pm$0.318 &4.070$\pm$0.302 &0.039$\pm$0.012 &2.987$\pm$0.829 \\ \hline  
kS01Pm1* &47 &-0.16 &16 &36.436$\pm$0.354 &35.382$\pm$0.098 &0.435$\pm$0.366 &8.707$\pm$0.718 &0.029$\pm$0.015 &2.839$\pm$1.442 \\
nS01Pm1* &\blue{43} &-0.19 &13 &27.413$\pm$0.291 &27.604$\pm$0.478 &0.552$\pm$0.378 &6.923$\pm$0.129 &0.036$\pm$0.014 &2.401$\pm$0.558 \\ 
qS01Pm1 &46 &-0.17 &18 &38.173$\pm$0.653 &36.291$\pm$0.830 &-1.033$\pm$0.829 &11.399$\pm$0.924 &0.039$\pm$0.011 &5.107$\pm$1.501 \\ \hline 
\blue{kS02Pm1 *} &66 &-0.23 &25 &70.218$\pm$6.340&65.830$\pm$6.636&0.717$\pm$0.298&32.987$\pm$3.260&0.020$\pm$0.009&4.164$\pm$1.529 \\ 
\blue{nS02Pm1 *} &62 &-0.25 &28 &59.060$\pm$3.234 &57.019$\pm$2.562 &0.390$\pm$0.554 &29.457$\pm$2.578 &0.020$\pm$0.011 &3.391$\pm$1.806\\ 
qS02Pm1 &65 &-0.24 &28 &70.091$\pm$0.825 &70.265$\pm$0.564 &-0.568$\pm$0.277 &42.854$\pm$0.917 &0.027$\pm$0.008 &5.468$\pm$1.317\\ \hline
\end{tabular}
\caption{Summary of the $\Pm$=1 
runs
with varying shear rate.}
\label{modelsPm1}
Note. Runs marked with `q',
`k' and `n'
have been analyzed with the quasi-kinematic TFM (QKTFM),
the kinematic version of
the CTFM (no main run), and
the
CTFM 
including the main run, respectively.
Boldfaced: runs
most compatible with \cite{ZB21}.
Runs marked with * indicate cases, where the magnetic background  turbulence is not 
statistically similar.
\end{table*}

\begin{table}[h]\caption{
Dynamo numbers for the runs with $\Pm$=1 and variable shear rate.}
\begin{center}
\begin{tabular}{@{\hspace{0.3mm}}lccccrrr@{\hspace{0.3mm}}} \hline \hline
Run           &$-\Sti$\phantom{a} & $\!\!\!\!\!{k_{z,\rm kin}}$ & $\!\!\!\!\!{k_{z,\rm sat}} $& $\!\!\!\!\!{\kB} $&$D_{\eta S}$ &$D_{\eta_{\rm rms} S}\!\!\!\!$& $D_{\alpha S}$\\    
                 &           &$\!\!\!\!\!{[k_1}]$              &$\!\!\!\!\!{[k_1]}$              &$\!\!\!\!\!{[k_1]}$&                     &                                             &\\ \hline \hline
nS00Pm1  & 0      &1 &1 &1 &9$\times 10^{-5}$        &0 &0 \\
qS00Pm1  & 0      &1 &1 &1 &2$\times 10^{-5}$        &0 &0 \\ \hline
kS005Pm1& 0.05 &1 &1 &1 &-0.0135  &0.2961 &0.7048 \\    
nS005Pm1& 0.05 &1 &1 &1 &0.0169   &0.3265 &0.6812 \\
qS005Pm1& 0.05 &1 &1 &1 &0.0585   &0.3350 &0.6829 \\ \hline
kS01Pm1  & 0.1   &1 &1 &1 &-0.0609  &0.4168 &0.8029 \\
nS01Pm1  & 0.1   &1 &1 &1 &-0.1870  &0.5908 &1.1218 \\
qS01Pm1  & 0.1   &1 &1 &1 &0.1339   &0.6988 &1.0811 \\  \hline
kS02Pm1  & 0.2   &1 &1 &1 &-0.0542  &0.3496  &0.4357\\
nS02Pm1  & 0.2   &1 &1 &1 &-0.0076  &0.0973  &0.0600\\
qS02Pm1  & 0.2   &1 &1 &1 &0.0401   &0.4317  &0.5277\\
\hline
 \end{tabular}\\[2mm] 
 \end{center}\label{dynamonumbersPm1}
Note. 
The dynamo numbers are calculated based on the
wavenumber $k_{z,\rm kin}$
of the fastest growing Fourier mode 
in the kinematic stage of the LSD.
The other wavenumbers 
are those of the mode dominating
in the saturation stage of the mean field, 
$k_{z,\rm sat}$, and those 
of the test fields, $\kB$. 
For the kinematic CTFM
measurements (label `k', no main run), $k_{z,\rm kin}$ and $k_{z,\rm sat}$
 have been taken from the corresponding 
CTFM run with main run (label `n'). 
\end{table}

\begin{figure}[h!]\begin{center}
\hspace*{-0.5cm}
\includegraphics[width=1.1\columnwidth]{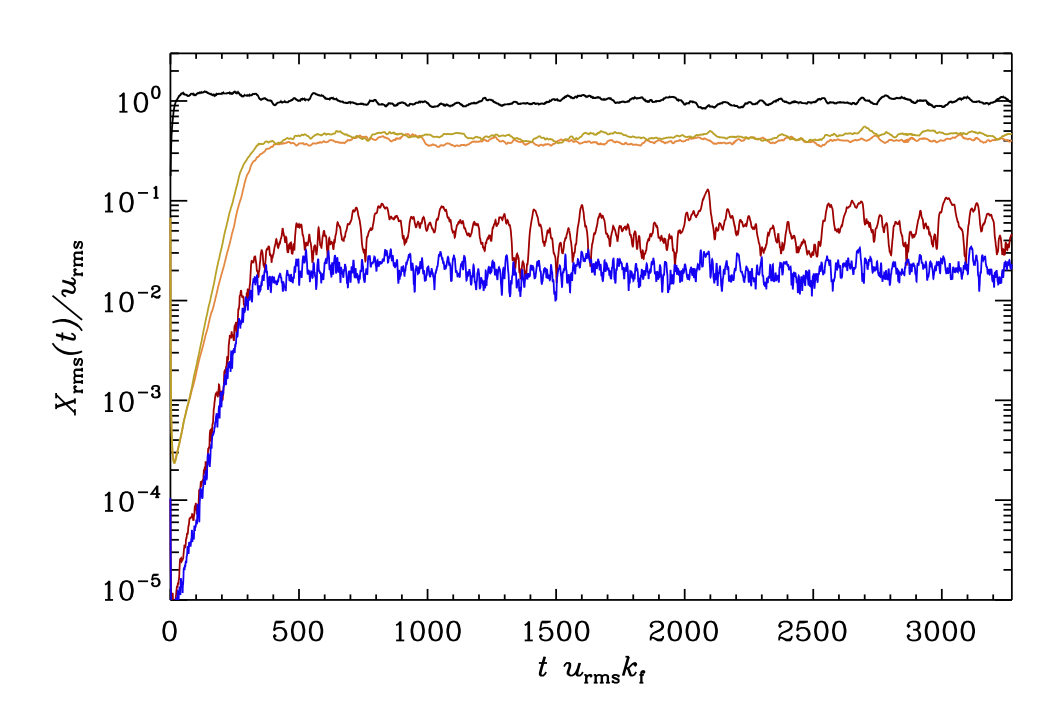}
\end{center}\vspace*{-0.5cm}\caption{
Volume-averaged rms values of the velocity ($\urmst$, black) and the total magnetic field from the main run ($\Brmst$, yellow), magnetic zero solution ($\brms^{(0)}(t)$, orange), 
rms values of the mean azimuthal ($\meanB_{y,\rm rms}$, red) and radial ($\meanB_{x,\rm rms}$, blue)
magnetic fields from Run~nS02Pm1, all normalized to $\urms$.
}\label{rmsPm1}\end{figure}

\begin{figure}[h!]\begin{center}
\hspace*{-0.4cm}
\includegraphics[width=1.1\columnwidth]{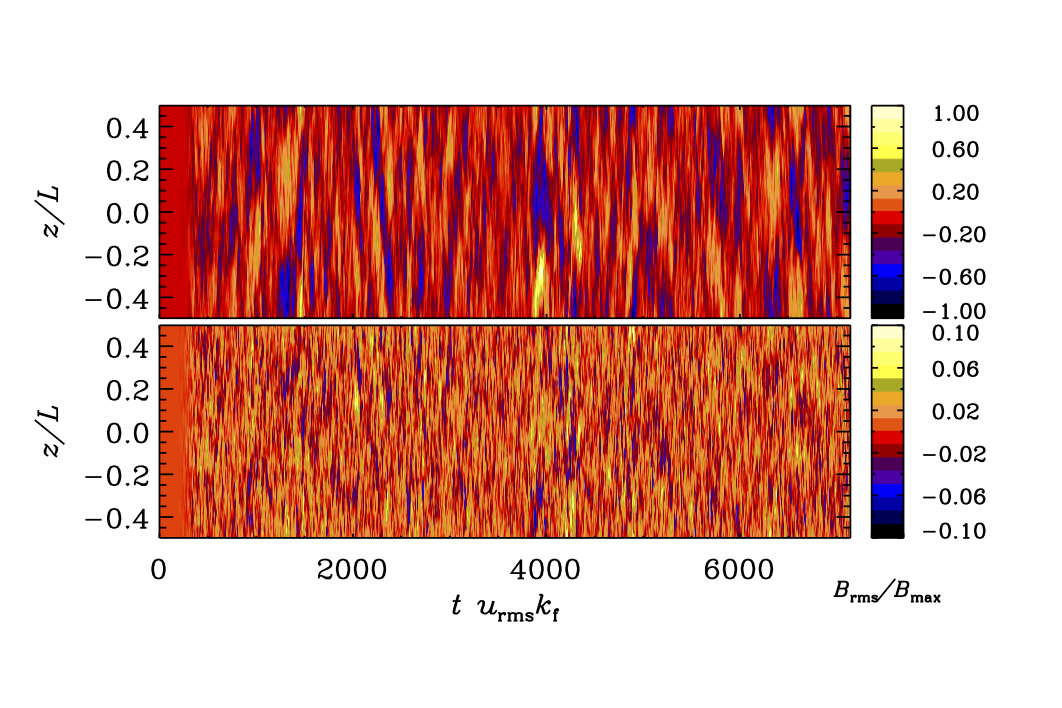}
\end{center}\vspace*{-0.8cm}\caption{
$zt$ diagrams 
of $\meanB_y$ (top)  and  $\meanB_x$ (bottom) from 
the main run of qS02Pm1. The main run of nS02Pm1 is identical up to some slight differences
due to the difference of time steps.
}\label{butPm1}\end{figure}

\subsubsection{\blue{Kinematic runs}}
\label{kinshear}
\blue{
We start by presenting the kinematic CTFM analysis of
some shear dynamo models.
We  stress
 that this method is generally valid and indispensable in studying the
possibility of  large-scale dynamo instabilities
with magnetic background turbulence  $\bbz$,
which is the main goal of this paper.
We vary $\Pm$ and $\ShK$, and study the scale dependence of the turbulent
transport coefficients by changing the vertical wavenumber of the test
fields, $k_B$.
We change 
$\Pm$
by keeping the magnetic resistivity $\eta$ fixed.
Hence, the larger $\Pm$, the smaller is $\Rey$.
The results are summarized in Table~\ref{modelskin}.
}

\blue{
In practice, we ran first only
the zero problems until saturation, and then forked 
new runs with the rest of the test problems
turned on.
The measurements 
would not be meaningful before saturation, but this strategy also 
accelerates the runs.
Despite the low Reynolds numbers, the systems are already prone to
chaotic behavior, and any small perturbation (such as the random seed
of the forcing being initialized differently,
or the time step changing slightly when 
$\kB$ is altered) can lead to different
small-scale dynamo solutions (for an extreme example,
compare the entries marked with a star in Table~\ref{modelskin}).
Hence, 
for investigating the scale dependence of the coefficients
it is desirable to choose runs that have as similar as possible 
SSD solutions.
A general observation 
is that the stronger
the SSD, 
that is, the
stronger the magnetic background turbulence, the
weaker are $\eta_{xx}$, $\eta_{yy}$, and $\eta_{xy}$, while $\eta_{yx}$
is always weak and consistent with zero within error bars.}

\blue{
We can also notice that 
the 
fluid
Reynolds number 
$\Rey$
is crucial for the magnitude of the $\eeta$ components and their fluctuations, 
all growing with $\Rey$,
which is clearly visible when comparing sets with $\Pm$=1 and 20 at
the weakest shear. In these, the SSD is weak, and the main effect must come from the flow:  
For  $\Pm=1$, the diagonal components and $\eta_{xy}$ are more than twice,
while $\eta_{yx}$ and $\eta_{\rm rms}$ an order of magnitude larger.
The magnitude of $\alpha_{\rm rms}/\etat \kf$ remains roughly constant,
but note that $\etat = \left(\eta_{xx} + \eta_{yy} \right)\!/2$.}

\blue{
Continuing with analyzing the sets with weakest shear, $\Sti=-0.05$, 
that is,
$-0.15<\ShK<-0.09$,
in the high $\Pm$ case, we find the diagonal $\eeta$
 components to be roughly equal, i.e. isotropic within error bars. 
The low $\Pm$ cases with $\kB/k_1=1$ show no anisotropy either, but this is not equally clear for the higher $k_B$ cases, 
which show weak anisotropy; this could be due to insufficient integration time 
though, as no significant anisotropy is expected in these  
weak shear runs without mean magnetic fields and stratification.}
\blue{
The high $\Pm$ runs show only weak scale dependence, insignificant within error bars, while the low $\Pm$ runs 
 show a clearly discernible one, such that the diagonal components are reduced when $k_B$ is increased. 
$\eta_{yx}$ is first positive, but turns to negative 
at the highest $k_B$;
yet
all values are
consistent with zero within error bars. The fluctuating quantities show no marked scale dependence at neither 
Prandtl number.}

\blue{
At moderate shear ($\Sti=-0.1$, resulting in $-0.30 <\ShK -0.16$), the diagonal
components show a weak anisotropy with both Prandtl numbers investigated. 
However, for $\Pm=1$, $\eta_{xx}$ is larger than $\eta_{yy}$,
while the opposite is true for $\Pm$=20.
$\eta_{yx}$ 
is negative, but consistent with zero within error bars for $\Pm$=20, while significantly positive for $\Pm$=1, for all $k_B$.
As per the scale dependence, the diagonal $\eeta$ components and
$\eta_{xy}$ are decreasing with $\kB$ for $\Pm=1$,
while $\eta_{yx}$ is constant. 
$\eta_{\rm rms}$ is decreasing  with $k_B$, too.
The trends in the set with $\Pm$=20 are just opposite for the diagonal
components and $\eta_{xy}$: they increase with $k_B$.
These differences must reflect the larger influence of shear on the flow and the stronger SSD generated in the case of $\Pm$=20.}

\blue{
At higher shear ($\Sti=-0.2$, resulting in $-0.60 < \ShK < -0.23$), the $\eeta$
anisotropy can be observed to get unified: with all the Prandtl numbers studied, 
$\eta_{xx}$ is systematically larger than $\eta_{yy}$
being statistically significant especially for $\Pm=20$.
The diagonal components exhibit only a weak scale dependence in all 
sets except $\Pm=1$, where a clear decrease as a function of $k_B$ is seen. 
$\eta_{yx}$ has a clear tendency of being positive or consistent with zero. 
$\eta_{xy}$ 
is the component showing the most prominent scale dependence 
with all $\Pm$ used: its magnitude decreases  
from $\kB/k_1=1$ to $\kB/k_1=3$ 
in all runs, 
although for $\kB/k_1=2$, one often finds an increased value in comparison to 
$\kB/k_1=1$.
Also  $\eta_{\rm rms}$ is decreasing as function 
of $k_B$, more prominently so the smaller $\Pm$ is. We also note that with 
high shear and $\kB/k_1=1$, $\eta_{xx}$ shows high-frequency
oscillations, which vanish at $\kB/k_1=3$.
}
 
\blue{
Finally, we have run one set with $\Sti=-0.3$, $\ShK \approx -0.86$, and $\Pm$=20. 
This set has the strongest SSD, but in comparison to the second largest shear rate, $\Sti=-0.2$,
the transport coefficients are no longer quenched strongly by it. Otherwise the anisotropy and scale dependence of the measured coefficients is very similar.
}
 
\blue{To summarize, the main findings in this section are: the SSD
quenches the $\eeta$ components up to a certain point, while more
vigorous kinetic turbulence, quantified by increasing $\Rey$, enhances their magnitude.
Scale dependence is evident only in runs with high enough $\Rey$.
Strong shear leads to anisotropy,
 with $\eta_{xx} > \eta_{yy}$.
Kinematic calculations show no evidence for negative values of $\eta_{yx}$ within the studied parameter regime.}
 
\subsubsection{Models with Prandtl number of unity}\label{prandtl}

We proceed by discussing runs with $\Pm$ of unity and forcing wavenumber \blue{$\kf/k_1=5$},
but the main run with the potential of LSD now included.
This choice is motivated by a recent study by \cite{ZB21}, who highlighted
weak-shear cases at low to moderate Reynolds numbers 
and $\Pm=1$ to give rise
to a negative $\eta_{yx}$, when measured with the QKTFM.
Without further analysis, such a result could be easily interpreted 
to be (i) contradictory to all previously published numerical results that 
have {\it not} reported negative $\eta_{yx}$ values in  incompressible, anelastic, or SMHD, 
and (ii) favorable for SC effect dynamos. 

As in \Sec{kinshear}, we start the measurements only after saturation of the zero problems
and
proceed 
until  the saturation of the evolving mean fields. 
Runs of this kind
are reported in our tables below
with names starting with `n' (nonlinear).
\blue{The runs starting with 'k' refer to the corresponding kinematic runs
presented in the previous section.}
We also perform QKTFM measurements for each run
(run names starting with `q').
\blue{In the `n' and `q' cases},
we compute the turbulent transport coefficients after the saturation    
of both types of dynamos (if present).

We begin by presenting results from two runs
without shear and rotation employing
the CTFM and QKTFM, respectively, and integrated over 
six thousand turnover times; 
see the first entries in Table~\ref{modelsPm1}.
For these setups, a SSD is present, resulting in 
initial exponential growth of the magnetic field, which then
saturates at around 400 turnover times. After that, the magnetic energy 
stays close to its average saturation value of
$\Lu \approx 0.25 \, \Rm$.
We see that both methods
give consistent results: the diagonal elements are isotropic, i.e., roughly of the same
magnitude within error bars, and the off-diagonal elements are consistent with zero.
These runs were integrated 
twice as long as any other run,
hence their
 error bars reflect the minimal level to be achieved with realistic computation times.
The agreement of the CTFM and QKTFM is not self-evident:
Given the weakness of the SSD,
 we interpret it as an indication of strong dominance of the contribution 
 $\overline{\uuz\times\bbB}$
to the mean EMF over $\overline{\uuB\times\bbz}$; see \Sec{sec:qktfm}.

By adding shear, we find that the
SSD is enhanced in terms of its saturation strength, as indicated by the increasing Lundquist number ($\Lu$) in Table~\ref{modelsPm1}. 
The growth rate also increases somewhat as function of the shear rate; for
a typical case see \Fig{butPm1} with the largest shear number in this set.
For the shear numbers covered, the 
rms strength of the total magnetic field 
reaches $\Lu \approx 0.42 \Rm$ at the highest shear.  
In all cases, the 
magnetic fields of the
zero problem and the main run 
are of similar strength;
correspondingly,
the mean 
fields reach maximally
a few percent of the equipartition strength
at the highest shear 
rate.
None of the constituents of the magnetic field show significant growth 
after the SSD has saturated; see \Fig{rmsPm1}.
Obviously, shear and turbulence have only the capability of generating short-lived 
large-scale patches in $\meanB_y$ (see \Fig{butPm1}), persisting only 
over a few tens to maximally a couple of hundred turnover times.
The timescale of persistence is even shorter in $\meanB_x$
(maximally a few tens). 
The wavenumber of these patches is \blue{$k_z/k_1=1$}, which is
\blue{ 
used both as the vertical wavenumber of the test fields in all the measurements, and}
to compute the dynamo numbers for these runs; see Table~\ref{dynamonumbersPm1}.

Differences in $\eta_{yx}$ between CTFM and QKTFM
start to emerge when shear is increased. 
\blue{As was reported in Sect.~\ref{kinshear}, the} 
kinematic CTFM 
\blue{(kCTFM)}
gives
consistently positive values \blue{for all values of $\Sti$, although for the weakest shear,
$\Sti=-0.05$, the error bars are too large to be certain of the positive sign.} 
\blue{Results in close agreement are} found
beyond kinematics.
This is expected as no clear LSD
is operational and thus the two versions of the CTFM have to agree.
The QKTFM, however, yields negative 
$\eta_{yx}$ 
within error bars with all 
shear numbers investigated. Hence, we can 
reproduce the results of \cite{ZB21} of negative $\eta_{yx}$ 
with the QKTFM, but the results of the CTFM do not lend support to them.

As per the other $\eeta$ components, we note that the \mbox{QKTFM} yields larger 
diagonal components
in comparison to the CTFM with main run,   
especially in the cases
with high shear. Both methods give positive values of 
$\eta_{xy}$, but the 
values retrieved with the QKTFM are larger than those with CTFM.

We note that in one case, namely with $\Sti=-0.1$, the background 
turbulence, \blue{both in terms of the flow and magnetic field zero solutions $\uuz$ and $\bbz$,} 
is statistically different between the kinematic and 
\blue{nCTFM}
runs.
\blue{We re-iterate that such differences can emerge even in mildly
turbulent flows, e.g., due to differences in the time step.}
Simultaneously, we observe a difference in the measured transport coefficients, such that the diagonal
components are larger by about 25\% \blue{in the kinematic run}.
Given the weakness of the mean field, we attribute these marked differences to the difference in the background turbulence. 
\blue{In Section~\ref{kinshear}, we saw that SSD can 
diminish the transport coefficients, while a higher level of kinetic
turbulence enhances them.
In the present case, both effects are present, given
the larger $\Rm$ and $\Lu$ in the kinematic run.
This result provides evidence for the enhancing effect by more vigorous kinetic
turbulence being stronger than the suppressing effect by the SSD in this shear regime.}
In the case $\Sti=-0.2$,
\blue{differences in $\Rm$ and $\Lu$ can again be observed, again indicative of the background turbulence being different
between the kinematic and the nonlinear runs.
However, 
these differences are smaller than for $\Sti=-0.1$,
which explains the weaker impact on the coefficients ($\approx$ 15\%),
but weak mean-field effects cannot be ruled out either.}

We also compute the dynamo numbers, following the procedure described earlier, 
and report them in Table~\ref{dynamonumbersPm1}.
We note that in the shearless case, $D_{\eta S}$ is ideally zero, but not in practice due to limited accuracy
of the measurements.
With all the shear rates investigated, 
the dynamo number
remain subcritical both 
with respect
to the coherent SC effect and the incoherent ones \citep[for a complete analysis, 
see][]{SMHD}. This is in perfect agreement with the 
observation of the mean fields to remain
weak with hardly any growth, except for a slight increase in  
$\meanB_y$.
This is most likely the reason why in the earlier study of \cite{BRRK08} employing the QKTFM,
which concentrated on analyzing the regime with generation of strong 
large-scale fields, no attention was paid to 
this parameter regime. We also regard it to be insignificant for the investigation
of dynamos in shear flows. 

\begin{table*}[h]
\caption{Summary of the models with variable $\Pm$ and fixed shear rate.}
\hspace*{-2cm}
\begin{tabular}{lrrrrccrrcc} \hline \hline
{Run}  &$\Pm$&$\Rm$  &$\ShK$ &$\Lu$ & $\eta_{xx}/\eta$ & $\eta_{yy}/\eta$ & $\eta_{yx}/\eta$\phantom{aaa} & $\eta_{xy}/\eta$\phantom{aaa} &$\alpha_{\rm rms}/\eta_t \kf$  &$\eta_{\rm rms}/\eta$\\ \hline \hline
kS02Pm5 &5 &16 &-0.48 &17 & 4.765$\pm$0.776 &4.576$\pm$0.919 &0.019$\pm$0.070 &4.993$\pm$0.208 &0.036$\pm$0.017&0.370$\pm$0.175\\
\blue{nS02Pm5*}      & 5 &16 &-0.50 &22 &4.235$\pm$0.181 &4.210$\pm$0.220  &0.173$\pm$0.016 &0.634$\pm$0.136 &0.028$\pm$0.012 &0.194$\pm$0.087  \\ 
\blue{qS02Pm5*}      & 5 &16 &-0.49 &18 &5.910$\pm$1.429 &6.034$\pm$1.388 &0.273$\pm$0.125 &1.994$\pm$1.447 &0.025$\pm$0.009 &0.381$\pm$0.315 \\  \hline 
\blue{kS02Pm10} &10 &42 & -0.47 &44 &13.203$\pm$0.467 &12.637$\pm$0.611 &0.139$\pm$0.053 &6.440$\pm$0.306 &0.027$\pm$0.011 &0.542$\pm$0.239 \\
nS02Pm10    &10&42 &-0.47 &48 &14.832$\pm$1.066 &14.923$\pm$1.042&0.501$\pm$0.035 &3.710$\pm$1.178 & 0.026$\pm$0.011 &0.659$\pm$0.246\\ 
qS02Pm10    &10&42 &-0.47 &49 &13.276$\pm$1.528&13.541$\pm$1.641 &0.400$\pm$0.077 &2.160$\pm$1.844&0.030$\pm$0.011 &0.631$\pm$0.376\\ \hline 
\blue{kS02Pm20}  &20&33 &-0.60 &44 &6.099$\pm$0.140&5.372$\pm$0.232&0.041$\pm$0.013&1.569$\pm$0.784&0.046$\pm$0.013&0.326$\pm$0.084\\  
\blue{nS02Pm20}   &20&33 &-0.60 &47&7.356$\pm$0.807&5.608$\pm$0.486&0.000$\pm$0.042&0.874$\pm$0.326&0.041$\pm$0.020&0.202$\pm$0.128\\ 
qS02Pm20    &20&33 &-0.61 &47 &5.696$\pm$0.462&6.008$\pm$0.528&0.033$\pm$0.021&-
1.017$\pm$0.533&0.045$\pm$0.011&0.228$\pm$0.094\\ \hline
 \end{tabular}\\[2mm]
\label{modelsPmvar}
Note. Conventions as in Table~\ref{modelsPm1}, \blue{except for the
star indicating different levels of mean fields present in the models, while the background turbulence is roughly similar.
As per the scale of the test fields, in the kinematic CTFM runs we use $\kB$ corresponding to the one seen in the main run during the growth phase of the magnetic field, while in the 
\blue{nCTFM}
ones, $\kB/k_1=1$ is used in all cases.} 
\end{table*}

\begin{table}[h]\caption{
Dynamo numbers for the models with variable $\Pm$ and fixed shear rate.}
 \hspace*{-0.9cm}
\begin{center}
\begin{tabular}{@{\hspace{0mm}}l@{\hspace{1.5mm}}rcccrrr@{\hspace{1mm}}} \hline \hline
Run  & $\Pm$  &$\!\!\!\!\!{k_{z,\rm kin}}$ & $\!\!\!\!\!{k_{z,\rm sat}} $& $\!\!\!\!\!{\kB} $&$D_{\eta S}$ &$D_{\eta_{\rm rms} S}\!\!\!\!$& $D_{\alpha S}$\\    
         &  &$\!\!\!\!\!{[k_1]}$ &$\!\!\!\!\!{[k_1]}$ &$\!\!\!\!\!{[k_1]}$ &&&\\ \hline \hline
\blue{kS02Pm5}      &5 &2  &1 &2 &-0.1137  &2.3003  &4.2421\\ 
nS02Pm5      &5 &2  &1 &1 &-1.2616  &1.4233  &2.9659 \\ 
qS02Pm5      &5 &2  &1 &1 &-1.1108   &1.5666 &2.4237\\ \hline
\blue{kS02Pm10}    &10&2 &1 &2 &-0.3556  &1.4004 &3.4350 \\  
nS02Pm10    &10&2 &1 &1 &-0.9868  &1.3086 &2.9487 \\ 
qS02Pm10    &10&2 &1 &1 &-0.7421  &1.5195 &3.5596\\ \hline
\blue{kS02Pm20}    &20&3 &1 &3 &-0.1981  &1.5581 &3.1317\\ 
nS02Pm20    &20&3 &1 &1 &0.0124   &0.8021 &2.6719\\
qS02Pm20    &20&3 &1 &1 &-0.1564 &1.0812  &3.2568\\
\hline
\end{tabular}
\end{center}
 \label{dynamonumbersPmvar}
Note. Conventions as in Table~\ref{dynamonumbersPm1}.
\end{table}

\subsubsection{Varying Prandtl number and moderate shear}

\begin{figure}\begin{center}
\hspace*{-0.4cm}
\includegraphics[width=1.1\columnwidth]{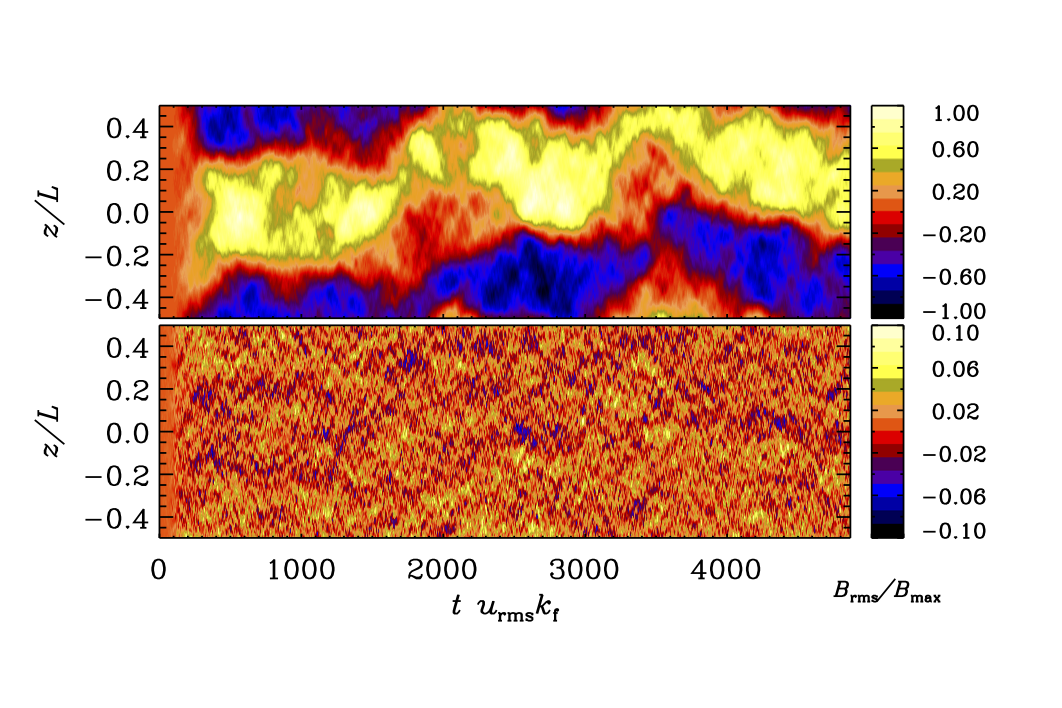}
\end{center}\vspace*{-0.8cm}\caption{
$zt$ diagrams 
of $\meanB_y$ (top)  and  $\meanB_x$ (bottom) from from 
the main run of qS02Pm20. The main run of nS02Pm20 is identical up to some slight differences
due to the difference of time steps.
}\label{butPm20}\end{figure}

\begin{figure}\begin{center}
\hspace*{-0.5cm}
\includegraphics[width=1.1\columnwidth]{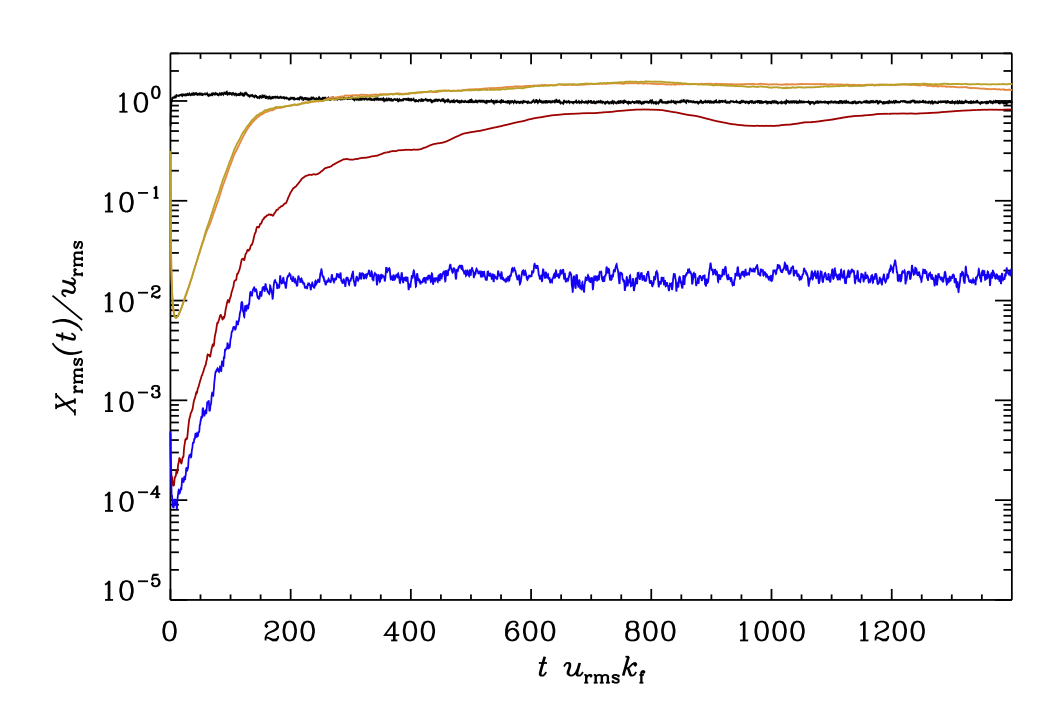}
\end{center}\vspace*{-0.5cm}\caption{
As \Fig{rmsPm1}, but for Run~nS02Pm20. 
}\label{rmsPm20}\end{figure}

Next we map out a part of the
parameter space where the generation of significant 
large-scale magnetic fields occurs.
One such regime can be found when
\blue{$\kf/k_1=10$ and $\Pm$ is} increased,
while keeping the shear number $\ShK$ at moderate values.
\blue{The results are summarized in Table~\ref{modelsPmvar}.}
For runs with $5\le\Pm\le20$, we find mean
magnetic field configurations
closely matching those of \citep[][]{BRRK08,SB15b}; see 
\Fig{butPm20}.
In contrast to the weak, incoherent, and short-lived patches seen in
\Fig{butPm1}, the mean azimuthal field grows to near equipartition;
see further \Fig{rmsPm20}.
The $k_z/k_1=1$ mode emerging in the nonlinear stage exhibits phase coherence
nearly throughout the whole 5000 turnover times of the run.
Likewise, $\meanB_x$ shows faint hints of the same pattern, 
but with a sign opposite to $\meanB_y$.

\begin{figure}\begin{center}
\hspace*{-0.5cm}
\includegraphics[width=1.1\columnwidth]{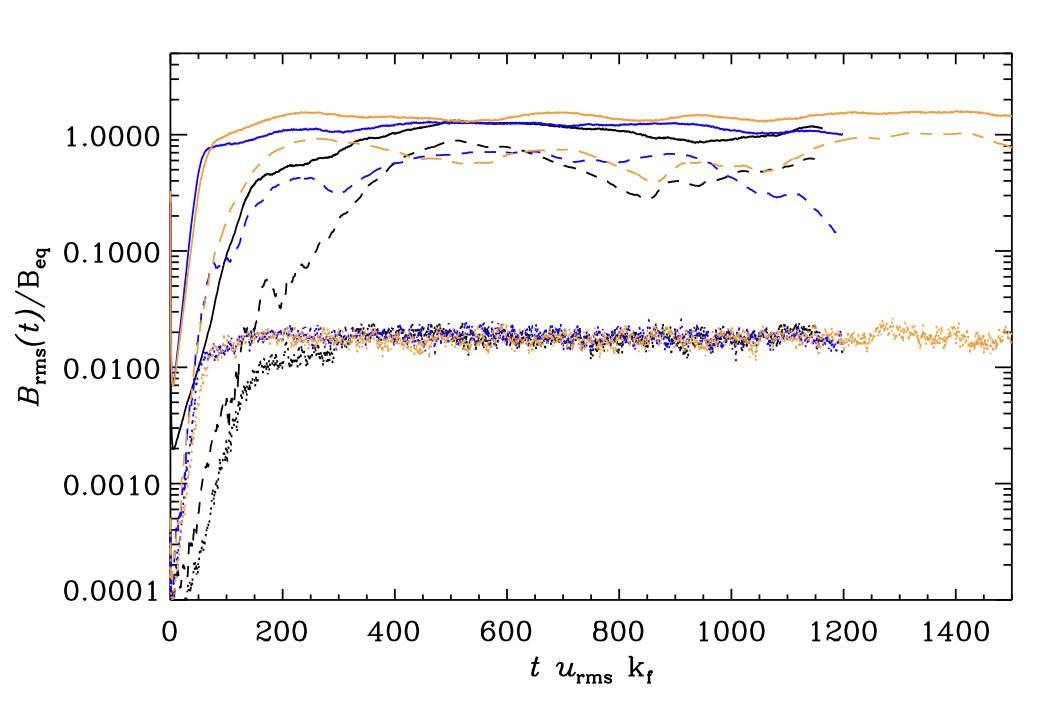}
\end{center}\vspace*{-0.5cm}\caption{
Volume-averaged rms values of the total magnetic field from the main run ($\Brmst$, solid), 
rms values of the
mean azimuthal ($\meanB_{y,\rm rms}$, dashed), and radial ($\meanB_{x,\rm rms}$, dotted)  fields from three different runs. 
Black: qS02Pm5, orange: qS02Pm20, blue: qS02Pm10.
}\label{mfs}\end{figure}

The properties of the obtained SSDs and LSDs are as follows: the mean azimuthal field grows 
always to near equipartition as can be seen from \Fig{mfs}.
The dynamo growth rates depend on $\Rm$, hence the SSD 
grows slowest for $\Pm=5$, while for $\Pm=10$ and $20$
nearly equally fast, yet with a higher growth rate.
For $\Pm=5$, the SSD saturates below equipartition,
but very close to it for the higher Prandtl numbers.
After saturation of the SSD, mainly $\meanB_y$
continues to grow.
Again, the highest $\Pm$ and $\Rm$ cases show the fastest growth, which
is, however, distinctly slower than that of the SSD.
For $\Pm=5$, the LSD grows the slowest, and, because its   
SSD's saturation strength was lower\footnote{
As we diagnose the mean field in terms of the rms values of 
$\meanB_{x,y}$, high vertical wavenumber modes enter.
Hence, if the SSD is strong 
and LSD is weak, these modes
will dominate, and
the growth of the 
low wavenumber modes
 (the actual mean field)
cannot necessarily be seen.
}, growth is seen also in 
$\meanB_x$.
Its growth rate, however, is different from that of $\meanB_y$,
which is somewhat atypical of ``standard'' dynamos and 
could be taken indicative of eigenmodes that consist of only one component; see \cite{Rheinhardt14,BC20} for examples.
However, for an LSD based on the coherent effects alone this can be ruled out here.
Eventually, the $\meanB_x$
components grow equally strong in all cases, while 
the saturation strength of $\meanB_y$ is 
the highest for the highest $\Pm$, but this component undergoes semi-regular oscillations in all the runs.
The total magnetic field saturation strength is also increasing with $\Pm$, see
the Lundquist numbers in Table~\ref{modelsPmvar}. 
Our data also reveals that
the saturation values of both 
$\big\langle{\bbz}^2\big\rangle$ and $\big\langle\meanBB^2\big\rangle$ grow with $\Pm$. 
This indicates that a possible reduction of $\big\langle\bb^2\big\rangle$ in comparison to $\big\langle{\bbz}^2\big\rangle$,
caused by the mean field (quenching), is overcompensated by $\big\langle\meanBB^2\big\rangle$
as the total magnetic energy is $\sim\big\langle \big(\bb + \meanBB\big)^2\big\rangle= \big\langle{\big(\bbz+\bbB\big)}^2\big\rangle+\big\langle\meanBB^2\big\rangle$. 

It is very difficult to disentangle the 
wavenumber of the preferentially growing
Fourier mode of the LSD, as the SSD `contaminates' the growth rates: all 
modes exhibit exponential growth
with the same growth rate as long as
LSD and SSD grow simultaneously
\citep[for a more detailed analysis of a similar system; see][]{MiikkaGPU},
and separating the growth rates of these two 
instabilities is impossible. Hence, we have to
rely on the following
means of separation:
We perform a dedicated
set of runs, where we first
suppress the mean magnetic field at each time step, 
while letting the SSD grow until saturation. After that we 
continue the simulations, but
with the mean fields allowed to grow from very small seeds.
Now only the eigenmodes of the LSD grow, so we can 
determine the fastest growing 
of them and extract its 
wavenumber 
$k_{z,{\rm kin}}$.
For $\Pm=5$ and 10 we obtain $k_{z}/k_{1}=2$, while
$k_{z,{\rm kin}}/k_{1}=3$ for $\Pm=20$.
As is evident from \Fig{butPm20}, in the saturated
stage, the mode of wavenumber unity takes over, 
thus $k_{z,{\rm sat}}/k_1=1$.
This happens in all  LSD--active simulations, as can be seen
from Table~\ref{dynamonumbersPmvar}.   
\blue{
In this section, we use the former (growth phase) wavenumber in the kinematic CTFM measurements 
and the latter (nonlinear phase) one in the nonlinear counterparts.
}

From Table~\ref{modelsPmvar}, we observe that with none of the 
employed TFMs the measured $\eta_{yx}$ is negative.
Hence, it is unlikely that these dynamos are driven by the 
coherent magnetic SC effect.
We observe that 
\blue{for the highest $\Pm$ investigated, the diagonal}
components of $\eeta$ 
\blue{get significantly anisotropic}
when measured with the CTFM,
\blue{such that  $\eta_{xx}$ is exceeding $\eta_{yy}$}.
Since this anisotropy is also 
recovered with the kinematic version,
it cannot solely be due to the mean magnetic field,
but must also reflect the growing influence of the shear, given that $\ShK$ is growing with $\Pm$.
Contrariwise,
we note that the QKTFM does not reveal this anisotropy at all.
Moreover, for \blue{the highest} $\Pm$, the two methods tend to return 
$\eta_{xy}$ with different sign -- still positive for CTFM, but negative for QKTFM.
\blue{QKTFM also shows 
opposite anisotropy
-- $\eta_{yy}$ exceeding 
$\eta_{xx}$ -- albeit
insignificant
within error bars.}

\blue{
Although the background turbulence 
of the kinematic and the nonlinear 
('n' and 'q') 
runs in this set 
 is satisfactorily
similar, the strength of the mean field 
may differ within
the latter ones,
depending on, e.g., how long the runs have been integrated,
or whether the mean-field was suppressed before the saturation of the SSD or not.
Hence, in Table~\ref{modelsPmvar}, we indicate the runs, where the mean field
strength is clearly different
with a star, see the $\Pm$=5 set.
Here, the 
\blue{kCTFM}
and 
\blue{nCTFM}
calculations yield very similar magnitudes of the diagonal $\eeta$
components, but QKTFM somewhat larger
ones, albeit with large error bars.
In the case of $\Pm=10$ and 20,
the diagonal $\eeta$ components 
from the nonlinear runs
exceed their 
kinematic counterparts.
The fluctuating coefficients are nearly always suppressed
in the nonlinear runs.
}

We can also observe that 
\blue{with the highest
$\Pm$, 
$\eta_{yx}$ is approaching 
zero with all of the methods used
\blue{while having been much larger and positive with lower $\Pm$}}.
This could be indicative of a tendency of 
$\eta_{yx}$ to change
sign as  $\Pm$ is increased further. 
We tried to investigate this 
regime with the CTFM, but observed
the test 
solutions
to get unstable, with super-exponential, 
likely unphysical growth. 
Accordingly, the measurements become unreliable. Preliminary 
results from the QKTFM, indeed indicate a sign change of 
$\eta_{yx}$ to negative, but without the 
possibility of properly utilizing the CTFM,
we leave this to be investigated in forthcoming work.

The kinematic dynamo numbers  listed in Table~\ref{dynamonumbersPmvar}, 
clearly predict positive growth rates for all $\Pm$,
as evidenced by the 1D mean-field dynamo model.
The 
\blue{nCTFM}
gives predictions
close to marginality, slightly subcritical for $\Pm=5$ and 10, and clearly critical for $\Pm=20$.
Those returned by the QKTFM do not predict dynamo action for $\Pm=5$,
but for larger $\Pm$, they are 
\blue{clearly supercritical, hence}
more consistent with the kinematic CTFM 
\blue{kCTFM}
measurements
\blue{with $\kB$ matching the vertical LSD wavenumber.}
In all cases in this set, the incoherent effects are
sufficient to explain dynamo action, with often slight, but
far from fatal inhibition by the coherent effect.
We note that for $\Pm=20$, the 
\blue{nCTFM}
yields positive dynamo numbers
$D_{\eta S}$ for the coherent SC effect, but these are clearly below the critical one;
moreover, the corresponding 
$\eta_{yx}$ 
values
turn out to be insignificant within error bars.

\subsubsection{Dependence on the shear rate}\label{shear}

Here,
we investigate the shear rate dependence of the transport coefficients and the LSD
at $\Pm=20$. We perform 
additional 
\blue{sets of} runs with shear rates $\Sti=-0.05, -0.1$, and $-0.3$. 
As per the efficiency of the LSD, we notice that the 
strongest mean azimuthal field $\meanB_y$
in terms of equipartition 
is obtained with $\Sti=-0.2$.
For that value, roughly 70\% of the magnetic
energy is in the mean field. 
For $\Sti=-0.1$ 
 the corresponding 
 fraction is 
30\%, and 
55\% for $\Sti=-0.3$.
Topology and coherence of the mean fields 
\blue{in the nonlinear stage}
do 
not change as a function of shear rate: in all runs, we see the 
dominance of a coherent 
$k_{z,\rm sat}/k_1=1$ 
mode. 
\blue{The modes growing in the kinematic stage have $k_{z,\rm kin}/k_1=2$ for the two lowest shear rates, and $k_{z,\rm kin}/k_1=3$ for the two higher ones.}
\blue{By comparing the 
\blue{kCTFM}
runs with those including the main run,
we can also observe that shear enhances the SSD:  in 
the case of the highest shear rate, the values of $\Lu$ are larger than in any other set}
with, at times, even superequipartition
\blue{$\bbz$}
\blue{for the two highest shear rates}. 
The presence of a more 
vigorous SSD, however, does not seem to boost the LSD
\blue{indefinitely with increasing shear, as the fraction of the energy in the mean field is lower in the $\Sti=-0.3$ case than in the $\Sti=-0.2$ one. }.

\blue{The correctness of the 
\blue{nCTFM}
with 
$\Lu$ as high as
for the highest shear rate,
$\Sti=-0.3$,
is no longer granted.
Hence, for this shear rate, we report measurements with the QKTFM and
\blue{kCTFM};
for all other shear rates we judge the method valid,
and report full sets of results}.
The 
diagonal components of $\eeta$ do not manifest marked anisotropy until 
$|\Sti| \geq 0.2$,
corresponding to 
$|\ShK| \gtrsim 0.5$. 
Then, the CTFM indicates 
$\eta_{xx}>\eta_{yy}$ 
within error bars, while the QKTFM  
rather tends to 
$\eta_{yy}>\eta_{xx}$, 
yet to be considered insignificant in view of the error bars.
\mbox{QKTFM} shows a sign change of $\eta_{xy}$, when
$|\Sti| \geq 0.1$, while the CTFM yields positive values.
The kinematic variant shows increasing values 
of $\eta_{xy}$
as function of shear rate, while the nonlinear variant indicates rather decreasing ones.
The most marked difference of the methods is seen for $\eta_{yx}$: the 
\blue{kCTFM}
yields negative, but insignificant values for weak shear, but then
significant positive values for strong shear.
The nonlinear version, in contrast, shows large positive values with weak
shear, and a much reduced value at the highest shear that this method
is applicable for.
The QKTFM is in rough agreement with the trends of the
\blue{nCTFM}.

The retrieved dynamo numbers 
(\Tab{dynamonumbersSvar})
indicate slightly subcritical incoherent dynamos for the lowest 
shear rate, but clearly supercritical incoherent dynamos for all other shear rates. The prediction 
for the coherent SC-effect dynamo is unfavorable, \blue{except for $\Sti=-0.1$},
where a positive, yet still subcritical 
$D_{\eta S}$
is obtained
\blue{with the 
kCTFM.}

\begin{table*}
\caption{Summary of the runs with varying shear rate and $\Pm=20$.}
\begin{center}
\begin{tabular}{@{\hspace{0mm}}lr@{\hspace{1mm}}rrrccrrcc@{\hspace{0mm}}} \hline \hline
Run  &$-S$\phantom{S}&$\Rm$  &$\ShK$ &$\Lu$ & $\eta_{xx}/\eta$ & $\eta_{yy}/\eta$ & $\eta_{yx}/\eta$\phantom{aaa} & $\eta_{xy}/\eta$\phantom{aaa} &$\alpha_{\rm rms}/\eta_t \kf$  &$\eta_{\rm rms}/\eta$\\ \hline \hline
kS005Pm20 &0.05&34 &-0.15 &16 &11.754$\pm$0.097&11.982$\pm$0.051 &0.024$\pm$0.080 &1.112$\pm$0.101 &0.039$\pm$0.021 &0.264$\pm$0.182  \\  
nS005Pm20&0.05 &34 &-0.15 &18&12.036$\pm$0.655 &12.191$\pm$0.630&0.133$\pm$0.061 &1.297$\pm$0.241 &0.043$\pm$0.013 &0.547$\pm$ 0.278\\
qS005Pm20&0.05 &33 &-0.15 &20 &11.116$\pm$2.280 &11.398$\pm$2.113 &0.135$\pm$0.089 &1.013$\pm$0.677 &0.046$\pm$0.020 &0.552$\pm$0.354 \\ \hline
kS01Pm20 &0.1 &32 &-0.30 &28 &7.678$\pm$0.283&8.151$\pm$0.231 &-0.092$\pm$0.064 &1.063$\pm$0.149 &0.036$\pm$0.020 &0.257$\pm$0.112  \\  
nS01Pm20  &0.1   &33 &-0.30 &28 &9.152$\pm$1.560 &9.506$\pm$1.551 &0.203$\pm$0.121 &1.236$\pm$0.878 &0.040$\pm$0.019 &0.345$\pm$0.233\\
qS01Pm20  &0.1   &32 &-0.31 &35 &6.696$\pm$0.309&7.097$\pm$0.308&0.041$\pm$0.011&-0.426$\pm$0.115&0.050$\pm$0.013&0.297$\pm$0.087 \\ \hline
kS02Pm20  &0.2 &33 &-0.60 &44 &6.099$\pm$0.140&5.372$\pm$0.232&0.041$\pm$0.013&1.569$\pm$0.784&0.046$\pm$0.013&0.326$\pm$0.084\\  
nS02Pm20    &0.2&33 &-0.60 &47&7.356$\pm$0.807&5.608$\pm$0.486&0.000$\pm$0.042&0.874$\pm$0.326&0.041$\pm$0.020&0.202$\pm$0.128\\ 
qS02Pm20    &0.2&33 &-0.61 &47 &5.696$\pm$0.462&6.008$\pm$0.528&0.033$\pm$0.021&-1.017$\pm$0.533&0.045$\pm$0.011&0.228$\pm$0.094\\ \hline
kS03Pm20 &0.3 &35 &-0.86 &56 &6.274$\pm$0.255 &5.359$\pm$0.161 &0.114$\pm$0.052 &5.511$\pm$0.480 &0.039$\pm$0.022 &0.241$\pm$0.129\\ 
qS03Pm20  &0.3   &34 &-0.87 &60 &5.813$\pm$0.311 &5.933$\pm$0.333 &0.055$\pm$0.022 &-0.693$\pm$0.695 &0.039$\pm$0.009 &0.188$\pm$0.062 \\ \hline
\end{tabular}
\end{center}
\label{modelsSvar}
Note. Conventions as in Table~\ref{modelsPm1}.
\end{table*}

\begin{table}[t!]\caption{
Dynamo numbers for the models with varying shear rate and $\Pm$=20.}
 \hspace*{-0.9cm}
\begin{center}
\begin{tabular}{@{\hspace{0mm}}lccccrrr@{\hspace{0mm}}} \hline \hline
Run  &$-S$& $\!\!\!\!\!{k_{z,\rm kin}}$ & $\!\!\!\!\!{k_{z,\rm sat}} $& $\!\!\!\!\!{\kB} $&$D_{\eta S}$ &$D_{\eta_{\rm rms} S}\!\!\!\!$& $D_{\alpha S}$\\    
         &       &$\!\!\!\!\!{[k_{1}}]$ &$\!\!\!\!\!{[k_{1}]}$ &$\!\!\!\!\!{[k_{1}]}$ &&&\\ \hline \hline
kS005Pm20&0.05  &2 &1 &2 & -0.0236&0.1952 &2.1389 \\
nS005Pm20&0.05  &2 &1 &1 &-0.0957 &0.3979 &2.0790 \\
qS005Pm20&0.05  &2 &1 &1 &-0.1110 &0.4594  &2.4033 \\ \hline
kS01Pm20  &0.1    &2 &1 &2 &0.2891 &0.8237 &4.5000 \\
nS01Pm20  &0.1    &2 &1 &1 &-0.4746 &0.8095 &4.3888 \\         
qS01Pm20  &0.1    &2  &1 &1&-0.1653 &1.1906 &6.2656\\ \hline
kS02Pm20    &20&3 &1 &3 &-0.1981  &1.5581 &3.1317\\ 
nS02Pm20    &20&3 &1 &1 &0.0124   &0.8021 &2.6719\\
qS02Pm20    &20&3 &1 &1 &-0.1564 &1.0812  &3.2568\\ \hline
kS03Pm20  &0.3    &3 &1 &3 &-0.7848 &1.4347 &2.6384 \\
qS03Pm20  &0.3    &3 &1 &1 &-0.3848 &1.3288 &3.4519 \\ \hline
 \end{tabular}
 \end{center}\label{dynamonumbersSvar}
Note. Conventions as in Table~\ref{dynamonumbersPm1}.
\end{table}

\section{Conclusions}

This work presents the compressible test-field method (CTFM) applicable to full MHD
with magnetic background turbulence.
We present an extensive set of tests using 2D 
velocity and magnetic fields of Roberts geometry, for which
 it has long been known
that the quasi-kinematic test-field method (QKTFM) completely
fails (giving even the wrong sign
of $\alpha$) 
when 
a 
(force-free)
magnetic background is forced
 \citep{RB10}.
  We find agreement 
in $\aalpha$
between the CTFM and the imposed-field method,
when the ratio of the (saturated) rms values of the mean field and the magnetic background turbulence,
 $\meanB_{\rm rms}/b^{(0)}_{\rm rms}$ is smaller than $\approx 7$.

\blue{Tests with the shear dynamo setup reveal agreement of two different
nonlinear flavors of the CTFM up to Lundquist numbers 
of $\meanBB$ of at least 25
while $\meanB_{\rm rms}/b^{(0)}_{\rm rms}$ is not exceeding $\approx 0.8$.
}
We also compare with the SMHD approach 
of our earlier study \citep{SMHD},
 neglecting the pressure gradient,
and find some mild discrepancies due
to the former omission of this term. 

We proceed by applying the CTFM to the case of shear dynamos, where our previous
study was limited to SMHD
with magnetic forcing, and was hence deemed inconclusive.
In this work, we use full MHD subject to 
kinetic non-helical forcing only
and moderate $\Rm$, yet
resulting in vigorous small-scale dynamo (SSD) action. 
\blue{We mostly concentrate on analyzing the kinematic CTFM 
results due to their general validity, while also presenting
results of nonlinear CTFM
(nCTFM)
within its validity range, and QKTFM ones for comparison.}
We largely
confirm the results of the earlier study, namely, 
the finding of
large-scale dynamos (LSD) excited
by the incoherent $\alpha$--shear effect, in the parameter regime of moderate shear numbers 
($\ShK \approx -0.3 \ldots -0.9$) and 
magnetic Prandtl numbers $\Pm = 5 \ldots 20$. With $\Pm=1$, where
\cite{ZB21} measured negative $\eta_{yx}$ (favorable for the coherent SC-effect dynamo) 
with the QKTFM, we find uninterestingly weak LSD, and the CTFM does not confirm
the QKTFM measurements. 

\blue{Parameter regimes studied in this work are limited to moderate $\Pm$ and $\ShK$.}
What
prevented us  
from extending
 our analysis  beyond these limits is the 
aforementioned
enhanced instability of
the test 
solutions
either in the presence of strong
mean flows or very strong magnetic fluctuations. Further studies in this regime might be 
enabled by using higher resolution and even smaller time steps,
greatly increasing the computational challenge, though. 
Another avenue for future research would be to assess the importance of
the density in the Lorentz force, where we have replaced it by a
constant reference value.
It is conceivable that 
density variability
becomes important at high Mach numbers,
which is a regime that has not yet received much attention; but
see \cite{RKB18} for specific predictions.

\begin{acknowledgements}
We acknowledge fruitful and inspiring discussions with 
Prof.\ Nishant Singh, Dr.\ Hongzhe Zhou, and
Dr.\ Jonathan Squire and Prof.\ Amitava Bhattacharjee in the Max Planck Princeton Center for Plasma Physics framework.
M.J.K.\ and M.R.\ acknowledge the support of the Academy of Finland
ReSoLVE Centre of Excellence (grant number 307411).
This project has received funding from the European Research Council (ERC)
under the European Union's Horizon 2020 research and innovation
program (Project UniSDyn, grant agreement n:o 818665).
A.B.\ acknowledges support through the Swedish Research Council (Vetenskapsr{\aa}det),
grant 2019-04234.

\end{acknowledgements}

\vspace{2mm}\noindent
{\large\em Software and Data Availability.} The source code used for
the simulations of this study, the {\sc Pencil Code} \citep{PC2020},
is freely available on \url{https://github.com/pencil-code/}.
The DOI of the code is https://doi.org/10.5281/zenodo.2315093 {\tt v2018.12.16}
\citep{axel_brandenburg_2018_2315093}.
The simulation setup and the corresponding data are freely available on
\url{http://www.nordita.org/~brandenb/projects/CompressibleTestfield/}.


\bibliography{mara}{}
\bibliographystyle{aasjournal}

\appendix
\section{\blue{Shear dynamo test experiments}} \label{sheartests}

\begin{figure}\begin{center}
\includegraphics[width=\textwidth]{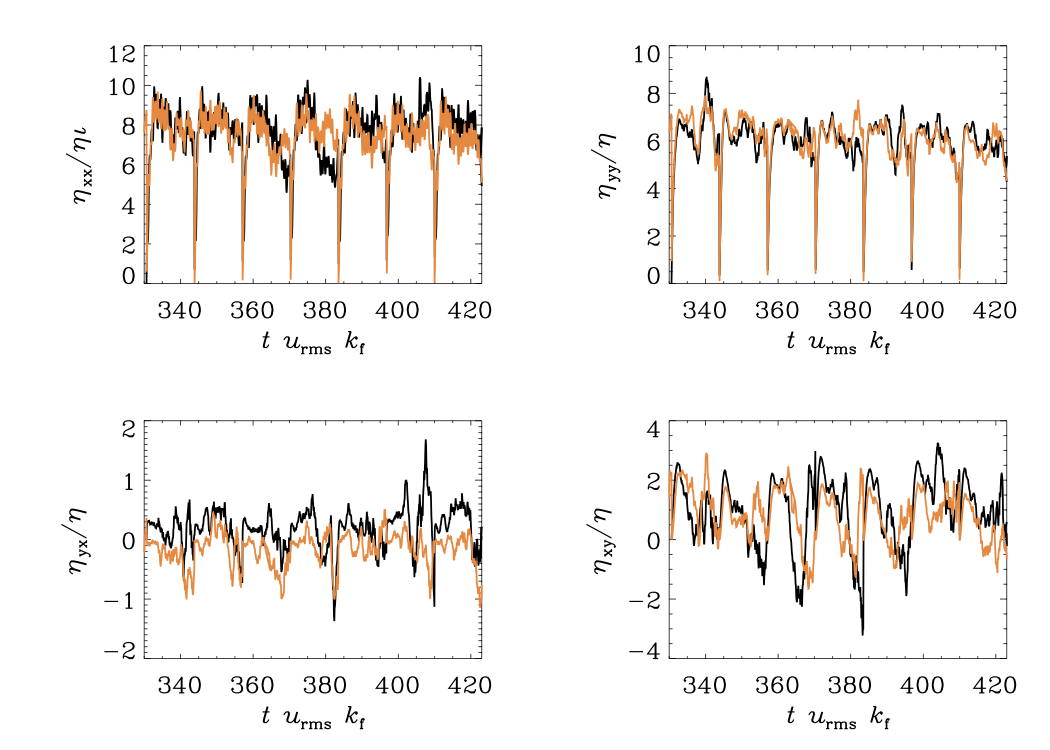}
\end{center}\vspace*{-0.8cm}\caption{
Comparison of the {\sf ju} (orange lines) and {\sf bb} (black lines)
flavors in the nonlinear stage for  $\Pm$=20 and $\Sti$=-0.2 
with $\Lu \approx$45 based on the total field strength, and $\approx$25 based
on the mean-field strength,
while $\meanB_{\rm rms}/b^{(0)}_{\rm rms}=0.78$
}\label{jubb}\end{figure}

\blue{
Given the results from the Roberts flow experiments, where the 
\blue{nCTFM}
failed to reproduce some of the $\aalpha$ coefficients for
imposed fields yielding Lundquist numbers above 10, it is important
to verify whether the nonlinear 
method is valid in
shear dynamo cases 
as the
generated mean fields at high shear and $\Pm$ reach a strong field
regime.
In case of the Roberts flow, we had 
a generally valid
method, namely the imposed
field one, to compare the CTFM results with, but for
determining $\eeta$ from a
shear dynamo
such an alternative does not exist.
Hence, we must rely on the comparison of the different flavors of the
CTFM, 
which may provide indication
for correctness, but no
definite proof.
We have re-run many of our shear dynamo runs with the {\sf 'bb'} flavor,
and show in Figure~\ref{jubb}  a 
comparison 
with  the {\sf ju} flavor
in the nonlinear
regime of a 
typical
case with a strong mean field.
We plot the time series of 
the $\eeta$ components
for seven resetting
intervals of the test problems.
The diagonal components from both methods agree very well, as can be
seen from the top row
of Figure~\ref{jubb}.
The agreement of the off-diagonal components (lower row) is somewhat
poorer, but still acceptable.
The $\eta_{yx}$ component from the {\sf bb} flavor shows a slight systematic offset to more positive
values, but as this component is small and its
time
average 
nearly always consistent with zero within error bars,
we conclude that this difference is not significant.
The agreement 
for $\eta_{xy}$ is again rather good.
}

\onecolumngrid
\section{\blue{Occasional correctness of the nonlinear method}}
\label{sec:correct}
\blue{
Consider the equations for $\aaB$, $\uuB$ and $\hB$ of the main run with imposed uniform field $\BB_0$
\EQA
\DDD^{A}\aaB&=&  \meanUU\times\bbB+\uu\times\BB_0+ \left(\uu\times\bbB + \uuB\times\bbz \right)'+\eta\nab^2\aaB,\label{daTnl}\\
\DDD^{U}\uuB  &=&- \nab \hB + \rho_{\rm ref}^{-1} \left[\jj\times\BB_0+\left(\jj\times\bbB + \jjB\times\bbz \right)' \right]\nonumber\\ && 
 - \meanUU\cdot\nab\uuB -  \uuB\cdot\nab\meanUU - \left(\uu\cdot\nab \uuB + \uuB\cdot\nab \uuz \right)'   \label{duTnl} \\
 &&+ \nu\!\left(\nab^2\uuB +\nab \nab\cdot\uuB/3\right) + 2\nu\left[ \meanSSSS \cdot\nab \hB  + \ssB \cdot\nab \meanH + \left({\sf s} \cdot\nab \hB + \ssB \cdot\nab \hz \right)' \right]/\cs^2 \nonumber \\
 \DDD \hB &=& - \meanUU\cdot \nab \hB - \uuB\cdot\nab \meanH  - \left(\uu\cdot\nab \hB + \uuB\cdot\nab \hz \right)'- \cs^2\nab\cdot\uuB ,\label{dhTnl}
\ENA
where the nonlinear terms are here written in the same way as described in \Sec{sec:nonlinear} for the {\sf ju} flavor of the test problems.
Accordingly, the mean EMF is expressed as $\meanEMF^{(B)}=\overline{\uu\times\bbB + \uuB\times\bbz}$.
Note, though, that here these writings represent equivalent rearrangements.
Now let us multiply the equations with a constant factor $g$ and redefine the variables as  $\aaB\coloneqq g\aaB$, ${\uuB}\coloneqq g\uuB$, ${\hB}\coloneqq g\hB$, and $\meanEMF^{(B)}\coloneqq g\meanEMF^{(B)}$. 
Of course, for $g\ne1$, no longer $\aaaa=\aaz+\aaB$ etc. holds.
We see that the system is now equivalent to that of flavor {\sf ju} of the test problems with the test field $\BB^{\rm T}$ set equal to the uniform field $g\BB_0$ (that is, $\BB^{\rm T}=\BB^{(1)}$ or $\BB^{(3)}$ with $k_B=0$ in \Eqs{tf1}{tf2} ).
Inverting the relation $\meanEEE_i^{(B)}= \alpha_{ij} B^{(T)}_j$ employing $\meanEMF^{(B)}$ derived from the test solution
must hence yield the same result as inverting $\meanEEE_i^{(B)}= \alpha_{ij} B_{0,j}$ with $\meanEMF^{(B)}$ derived from the main run (that is, employing the imposed field method).
For any other flavor, the same reasoning can be put forth, so they have to yield identical results.
If $\BB_0$ is, say, in $x$ direction, the CTFM 
with $\kB=0$
yields thus the correct $\alpha_{ix}$
for arbitrary strengths of the imposed field and hence the nonlinearity. However, $\alpha_{iy}$ cannot correctly be determined as only one of the test fields can be set 
proportional to $\BB_0$.}

\end{document}